\documentclass[twocolumn,prb,groupedaddress]{revtex4-1}

\usepackage[utf8]{inputenc}
\usepackage[english]{babel}
\usepackage[T1]{fontenc}
\usepackage{lmodern}
\usepackage{graphicx}
\usepackage{amsmath}
\usepackage{amsfonts}
\usepackage{amssymb}
\usepackage{microtype}
\usepackage{braket}
\usepackage{caption}
\usepackage{subcaption}
\usepackage{kantlipsum}
\usepackage[breaklinks=true,colorlinks, citecolor=blue]{hyperref}

\newcommand{\bi}{\bibitem}
\newcommand{\cc}{\captionsetup{justification=raggedright,singlelinecheck=false}}

\newcommand{\ct}{\cite}

\newcommand{\beq}{\begin{eqnarray}}
	\newcommand{\eeq}{\eeq{eqnarray}}

\begin{document}

\begin{@twocolumnfalse}

\title{Unitary preparation of many body Chern insulators: Adiabatic bulk boundary correspondence}
	\author{Souvik Bandyopadhyay} 
	\email{souvik@iitk.ac.in}
	\author{Amit Dutta}
	\email{dutta@iitk.ac.in}
	\affiliation{Department of Physics, Indian Institute of Technology Kanpur, Kanpur 208016, India}

	\begin{abstract}

{We approach the long-standing problem of preparing an out-of-equilibrium many-body Chern insulator (CI) and associated bulk-boundary
correspondence unitarily. Herein, this is addressed
by constructing a dynamical many-body  Chern invariant exploiting the property of the bulk macroscopic electric polarisation (Resta polarisation) of the CI. This Chern invariant defined from observable correlations is also established to topologically classify many body Chern states in equilibrium. The non-equilibrium behavior of the invariant is probed by ramping the paradigmatic Haldane model
of graphene from its  trivial to the topological phase. 
We show that a non-linear ramp may work more  efficiently in approaching the topological state, thereby establishing the existence of an optimal topological state preparation.
Furthermore,  to ensure the  {\it near} adiabatic dynamics across the quantum critical point, we propose a novel counter-diabatic scheme.
The topological nature of the prepared state is firmly established by {observing} an emerging $U(1)$ topological charge. We also compute the edge current in the time evolved state of the system under a semi-periodic boundary condition and clearly establish
an adiabatic bulk-boundary correspondence which firmly ensconces the validity of the many-body invariant. }

		
	\end{abstract}
	\maketitle
\end{@twocolumnfalse}

\section{Introduction}
There has been a recent upsurge in theoretical \ct{kitaev01,kane05,bernevig06,fu08,zhang08,sato09,sau10a,sau10b,lutchyn10,oreg10,moore10,shen12,bernevig13,haldane83,wen95,kitaev03,kitaev06,levin06} and   experimental \ct{mourik12,rokhinson12,deng12,das12,churchill13,finck13,alicea12,leijnse12,beenakker13,stanescu13} studies probing the generation and manipulation of topological phases of  many body quantum systems.  
Such topological phases are characterized  and distinguished by different  quantised values of a topological invariant which serves as a non-local order parameter. 
Distinct topological phases in thermodynamically large systems, separated by a quantum critical point (QCP) \ct{sachdev10,dutta15} exhibit strong robustness against external local perturbations. 
and thus, promise exciting new possibilities 
in understanding many body quantum phases stable under experimental situations with possible
for applications \ct{kitaev16,pachos17,damski11,nag12}.
 in topological quantum computation \ct{kitaev16,pachos17} and controlling dechorence \ct{damski11,nag12}.
The physical manifestation of `topology'  in symmetry protected topological  insulators (SPTs)  (see \ct{moore10,shen12,bernevig13}, for review) and Chern insulators (CIs) \ct{haldane83}  is rendered in the form of topologically protected {\it boundary-localised} zero energy states  when the bulk system is topologically non-trivial according to the bulk-boundary correspondence.
Although the equilibrium topology of  non-interacting quantum many body {systems is well understood, characterising the topology of systems 
which are driven 
out of-equilibrium \ct{oka09,bermudez09,kitagawa11,lindner11,cayssol13,rudner13,patel13,thakurathi13,kundu13,rajak14,balseiro14,mitra15,gil16,budich16,utso171,vajna15,utso_haldane_17,heyl13, sharma16,heyl18}  
still remain a challenging task.

Dynamically engineering a non-equilibrium  topological system is a two-pronged process:
a)  dynamical generation of a topological Hamiltonian \ct{oka09,kitagawa11}; 
 b) preparation of the system in  a topologically non-trivial dynamical state, e.g., in the ground state of the effective topological Hamiltonian, which
 is  relatively difficult. {Despite several works \ct{foster13,foster14,rigol15,cooper15,utso17,sougato18,ginley18,souvik90,souvik191,verresen20,pastori20,hu20}, the topological characterization of
 out-of-equilibrium systems exhibiting a bulk-boundary correspondence, 
is still lacking.}
The dynamical topological invariant has been recently studied in out-of-equilibrium one dimensional (1D) topological system\ct{foster13,foster14,ginley18,souvik90,souvik191}. 
More importantly,  in Ref. [\onlinecite{souvik191}]
using  a periodic driving scheme with a linearly ramped amplitude, 
a stroboscopic  "out-of-equilibrium" bulk-boundary correspondence has been established 
for one-dimensional (1D) topological systems. Therein we establish that it is indeed possible to adiabatically deform a topological system and melt it in a different topological {state} without ever crossing a critical point in 1D SPTs. A similar
study has also been reported in bosonic topological phases through the introduction of interaction with a super-lattice \ct{mortuk17}. These studies establish that certain topological phases can indeed be continuously connected by expanding the accessible Hilbert space of the system or by breaking the protecting symmetries.\\

Interestingly, for two-dimensional (2D) CI systems (e.g., Haldane model \ct{haldane83}), a {\textit no-go theorem} has been postulated \ct{rigol15}, which states  that  the initial bulk  topology of the  model characterised by a dynamical Chern number (CN), must not change under a smooth unitary transformation in a thermodynamically large system.
Nevertheless, following an adiabatic  {quench}, 
the edge current in considerably large systems eventually thermalizes to a value corresponding to the topology of the post-quench Hamiltonian \ct{rigol15,cooper15,utso17,sougato18}; thereby, implying the absence of an out-of-equilibrium bulk-boundary correspondence with respect to a topological index of the translationally invariant system. This creates a paradox on the existence of any exclusive bulk topological origin of the post-quench edge current in CIs,{and hence there exists  a lacuna in the holistic characterisation of the out-of-equilibrium bulk topology of a CI.}\\
 
{We precisely address this issue by constructing a many-body Chern invariant which can be defined through observable correlators. This will thus allow an observable based study of out-of equilibrium topology, including its thermalization properties similar to the edge current.
We propose a generalised invariant using the the bulk Macroscopic electric polarisation (MEP) (Resta Polarisation) \ct{resta98} in the topological phase \ct{vanderbilt09} {which is a novel approach in itself} and also extending it to a non-equilibrium scenario. We further show that this generalised CN is allowed to vary dynamically
and may approach an integral value when the system is ramped from the non-topological to the topological phase in a near adiabatic fashion. {This is}  illustrated through the the Haldane model \ct{haldane83} {considering both linear and non-linear quenching protocols.} Although it is not possible to initiate a topological phase transition without closing the bulk gap, we approach the topological state through dynamics in finite size systems by approaching large system sizes, which is a realistic direction of approach experimentally. Recently, there have been studies that have probed the possibility to prepare topological states without going through a gapless point by the introduction of super-lattices \ct{mortuk17} and also a dissipative approach to many-body topological steady states \ct{souvik202}. {Moreover, in the case of the non-linear quenching, we find out an optimal rate that facilitates an efficient generation of the topological state.}
{To approach
adiabaticity feasibly in a large system, we also propose a counter-diabatic (CD) protocol to suppress otherwise
inevitable excitations  {in passage} through the minimum spectral gap.}  Furthermore, we explicitly demonstrate the dynamical occupation of topological edge states and thus establish an adiabatic bulk-boundary correspondence with respect to the many-body Chern number in a clean CI.\\

Furthermore, the many-body measurable {topological invariant} we propose would generically {approach integral values} under unitary dynamics as non-equilibrium generation of excitations is progressivly suppressed in sufficiently large systems. Recently the method  has also turned out to be useful for a dissipative preparation of many-body Floquet Chern insulators \ct{souvik202}. It has also been reported lately that the many-body invariant we propose can indeed be directly measured in equilibrium through randomized measurements \ct{cian20}.}\\

{ The paper is organised in the following fashion: In Sec. \ref{sec_MEP},  we introduce the notion of a gerenalised CN using the property of the MEP both in equilibrium and
non-equilibrium situations and establish its topological properties. In Sec. \ref{sec_illustration}, on the other hand, we illustrate the concepts presented in Sec. \ref{sec_MEP} conidering
the linear as well as non-linear quenching of the Semenoff mass of the Haldane model of graphene. In this section, we also discuss the CD protocol at length emphasising its
significance and short-comings. In Section \ref{sec_bulk_boundary}, we calculate the edge current in the final evolved state and show the existence of an adiabatic bulk-boundary correspondence. Concluding comments are presented in Sec. \ref{sec_outlook}. Further, we have added five appendices to complement the discussion in the main text. } {We note, at the outset, that everywhere in this paper, we have used $\hbar=k_B=1$ such that all quantities and observables are specified in natural units.}

\section{MEP and Chern topology}
\label{sec_MEP}
 {
The macroscopic dipole polarisation vector \ct{resta98,vanderbilt09} of a band insulator in the directions of the lattice basis vectors ${\bf \hat{a}_i}$ (see Appendix \ref{sec_App_MEP} for an elaborate discussion), is defined as  $\vec{P}=\sum\limits_{i}P_{\hat{i}}{\bf \hat{a}_i}$. 
Here $P_{\hat{i}}=\left<\hat{X}_i\right>$, $\hat{X}_i$ being the many-body position operator and the expectation is taken over occupied single-particle states. The operator $\hat{X}_i=\sum\limits_{n}x_{i}^n\hat{a}^{\dagger}_n\hat{a}_n$ is the many-body position operator where $x_i^n$ denotes the coordinate of an atom at the $n^{th}$ site along the $i^{th}$ lattice direction with $a_n^{\dagger}$ being the corresponding  fermionic creation operator at that site. The expectation is taken over a fermionic many body state. The  {momentum} translation operator in the $i^{th}$ direction under periodic boundary conditions,
\begin{equation}
\hat{T_{i}}(\delta_i)=e^{i\delta_{i}\hat{X}_{i}},
\end{equation}
where we choose $\delta_i=2\pi/L_i$, $L_i$ being the dimension of the system in the $i^{th}$ direction.

Under periodic boundary conditions, the above definition may be compactified as,
$P_i={\rm Im}\ln\left<\hat{T}_i\right>$, where $\hat{T}_i$ is the momentum translation operator.
It then follows, that in the thermodynamic limit (see \onlinecite{vanderbilt09}),
\begin{equation}\label{eq:pure_pol_1}
P_i\left[\vec{k}_0\right]=\sum\limits_{\alpha}{\rm Im}\int_{BZ\left[\vec{k}_0\right]}\braket{\psi_{k,\alpha}|\partial_{k_i}|\psi_{k,\alpha}}dk_{1}dk_{2},
\end{equation}
where the {Brillouin zone} (BZ) is spanned by the reciprocal lattice vectors $\vec{b}_1$ and $\vec{b}_2$ such that,
$\vec{k}=k_1\vec{b}_1+k_2\vec{b}_2$, where, $k_1$, $k_2\in[0,1]$.
$\vec{k}_0$ having components $k_{01}$ and $k_{02}$ along the directions $\vec{b}_1$ and $\vec{b}_1$, is chosen to be the origin of the BZ. Here, $\ket{\psi_{k,\alpha}}$'s are the respective {occupied} single particle bands labelled by $``\alpha"$.}\\

 {Unlike in a 1D system, the MEP is not itself a topological quantity in 2D. However, in the topological phase, the macroscopic dipole polarisation is not a well-behaved quantity due to the absence of localised Wannier functions or equivalently, the presence of conducting edge states. Rather, for every adiabatic shift in the origin of the BZ, the MEP vector changes proportionally to the CN (Note that an adiabatic shift implies that  the occupation of each band remain invariant). For a band insulator, it can be shown that for an infinitesimal shift in {$\delta \vec{k}_0$} in the origin $\vec{k}_0$ of the Brillouin zone,
\begin{equation}
\delta P_i[\vec{k}_0]=P_i[\vec{k}_0+\delta \vec{k}_0]-P_i[\vec{k}_0]=2\pi\epsilon_{ij}\delta k_{0j} {\cal C},
\end{equation}
where $\mathcal{C}$ is the CN {and $\epsilon_{ij}$ is the antisymmetric tensor}.} \\

We utilize this non-uniqueness of the MEP in the topological phase \ct{vanderbilt09}, to conjecture a generalised CN as,
\begin{equation}\label{eq:chern_diss}
\mathcal{C}=\epsilon_{ij}\frac{\delta P_i[\vec{k}_0]}{2\pi\delta k_{0j}}.
\end{equation}
 {In a half-filled system at equilibrium, the Chern number reduces perfectly to the conventional CN \ct{shen12}}.\\

\subsection{Generalised CN and unitary dynamics}
 We start from an initial eigenstate $\ket{\psi(0)}$ of a CI
  in the non-topological phase having ${\mathcal C}=0$ , 
 which is subjected {to} an arbitrary unitary  time dependent drive.
To define the out-of-equilibrium CN, 
}
we extend the quantity defined in Eq.~\eqref{eq:pure_pol_1}, 
to a weighted average over the instantaneous {bands of single-particle states},
\begin{equation}\label{eq:ct_def}
\tilde{P}_i=\sum\limits_{\alpha}{\rm Im}\int_{BZ}dk_{1}dk_{2}n_{\alpha}^k(t)A^k_i(\ket{\phi_{k\alpha}(t)}).
\end{equation}
Here, $A^k_i(\ket{\phi_{k\alpha}(t)})=\braket{\phi_{k\alpha}(t)|\partial_{k_i}|\phi_{k\alpha}(t)}$ is the $U(1)$  {gauge connection on the single-particle eigenstate $\ket{\phi_{k\alpha}(t)}$ of the instantaneous Hamiltonian}
and the weights $n_{\alpha}^k(t)$ are {the time dependent population of the the instantaneous band `$\alpha$' as a function of momenta $k$},
{ i.e.,
\begin{equation}
n^k_{\alpha}(t)=\braket{\psi_k(t)|c^{\dagger}_{k\alpha}(t)c_{k\alpha}(t)|\psi_{k}(t)}.
\end{equation}
$n^k_{\alpha}(t)$ is the weighted average of the electric polarisation in each band of the time-evolved Hamiltonian $H(t)$; the weights being precisely the time dependent population of each band. This  in turn is manifested in the instantaneous current as we have thoroughly discussed in Appendix \ref{App_current} which also provides the motivation
behind defining the dynamical CN.}

We now proceed to define the dynamical CN as the change in the quantity $\tilde{P}_i$ {corresponding to} a shift $\delta \vec{k}_0$ in the BZ origin. This leads to the time-dependent CN,
\begin{equation}\label{eq:pure_chern_I}
\mathcal{C}^U(t)\propto 
\delta \tilde{P}_1[\vec{k}_0]=-\delta k_{02}\int_{k_{02}}^{k_{02}+1}dk_{2}\partial_{k_2}\beta(k_2,t),
\end{equation} 
\begin{equation}\label{eq:pure_beta}
{\rm where}, \beta(k_2,t)=-{\rm Im}\int_{k_{01}}^{k_{01}+1}dk_1 \tilde{A}^k_1(\ket{\phi_{k\alpha}(t)}),
\end{equation}
{where $\tilde{A}_i(t)= \sum\limits_{\alpha}n^k_{\alpha}(t)\langle{\phi_{k\alpha}(t)|\partial_{k_i}|\phi_{k\alpha}(t)}\rangle.$}\\

The quantity $\mathcal{C}^U$ defined in Eq.~\eqref{eq:pure_chern_I} is invariant under a local $U(1)$ gauge transformation owing to the non-interacting nature of the systems studied in this context(as elaborated in Appendix.~\ref{sec_App_MEP}). 
At {\it equilibrium}, when any one of the bands is completely filled, the quantity $\tilde{P}_i$, reduces to the total MEP of the occupied band. In this situation, the CN defined in Eq.~\eqref{eq:pure_chern_I} simply detects a branch change of the function $\beta(k_2)$ in the closed $\mathcal{S}^1$ interval $k_2\in[0,1]\equiv I$; which equivalently counts 
the winding of $\beta(k_2)$ along $k_2$ \ct{vanderbilt09}. {This implies that the existence of a branch singularity in the map $k_2\in[0,1]\rightarrow\beta(k_2)$, signals the Chern non-triviality of the system. In the following, we shall elaborately discuss different aspects concerning the topological nature of the dynamical CN defined above.}\\

Firstly, let us focus on the equilibrium topological characterization: the function $\beta (k_2)$ as described in the manuscript is nothing but a uni-directional Berry phase along one of the periodic directions $\mathcal{S}_a^1\equiv k_1\in[0,1]$ and defined at each point of the $\mathcal{S}^1_b$ interval $k_2\in[0,1]$. This decomposition into two circles $\mathcal{S}^1_a$ and $\mathcal{S}^1_b$ is possible because the Brillouin zone (BZ) forms a 2-tori $T^2$ which is topologically equivalent to,
\begin{equation}
	T^2\equiv \mathcal{S}^1_a\times \mathcal{S}^1_b.
\end{equation}
As shown in Eq.~(2) of the manuscript, the shift in the polarisation is directly proportional to the Chern number which is a $\mathbb{Z}$ topological invariant. Equivalently, a branch change of the function $\beta(k_2)$ in equilibrium at the ends of the BZ in a topological phase, immediately results in the non-uniqueness of the polarisation. We however observe that it is not essential for the branch singularity to occur at the end points of the BZ. In fact, the branch singularity of $\beta(k_2)$ at any point $k_2^*\in[0,1]$ reflects the topology of the system (as illustrated below). This is because the invariant $C^U$ defined in the manuscript simply provides with a homotopy classification of the map $\kappa_2\in\mathcal{S}_b^1\rightarrow\beta(k_2)\in\mathcal{S}^1$. In fact, the invariant $C^U$ reflects the integer winding of the function $\beta(k_2)$ as $k_2\in[0,1]$ which in turn is bound to be integer quantised as the fundamental homotopy group of the map $\mathcal{S}_b^1\rightarrow\mathcal{S}^1$ is $\pi_1(\mathcal{S}^1)\equiv\mathbb{Z}$.\\

{By fixing a gauge, such that $\beta(k_2)$ remains continuous for all $I: k_2\in[0,1]$ in a topological  phase, the function $\beta(k_2)$ exhibits a branch change proportional to the CN, at the endpoints of the BZ, i.e.
	$C\propto\beta(k_{02})-\beta(k_{02}+1)$ as in Ref.~\ct{vanderbilt09}.}

To elaborate, choosing a smooth gauge in $I$ ensures that the derivative $\frac{d\beta(k_2)}{dk_2}$ is well defined in the interval and its integration over the $\mathcal{S}^1\equiv I:k_2\in[0,1]$,
\begin{equation}\label{eq:1}
\frac{1}{2\pi}\int_0^{1}\frac{d\beta(k_2)}{dk_2}dk_2=\frac{1}{2\pi}\left[\beta(1)-\beta(0)\right]=-\Delta
\end{equation}
simplify reduces to the difference between the $\beta$ function evaluated at the "end-points" of interval $I$. Owing to the single-valuedness of the wavefunction at $k_2=0$ and $k_2=1$, this jump '$-\Delta$' is simply the integer quantised Chern number. Hence, a non-zero Chern index in this case implies a branch change of the map $B:\kappa_2\rightarrow\beta(k_2)$ after a complete rotation in $k_2\in \mathcal{S}^1$. \\
Now, since, the interval $I$ forms a complete circle $\mathcal{S}^1$, the occurrence of the branch change at any other point $k_2^*$ can also 
\begin{figure*}
\centering
\begin{subfigure}{0.45\textwidth}
\centering
\includegraphics[width=\columnwidth]{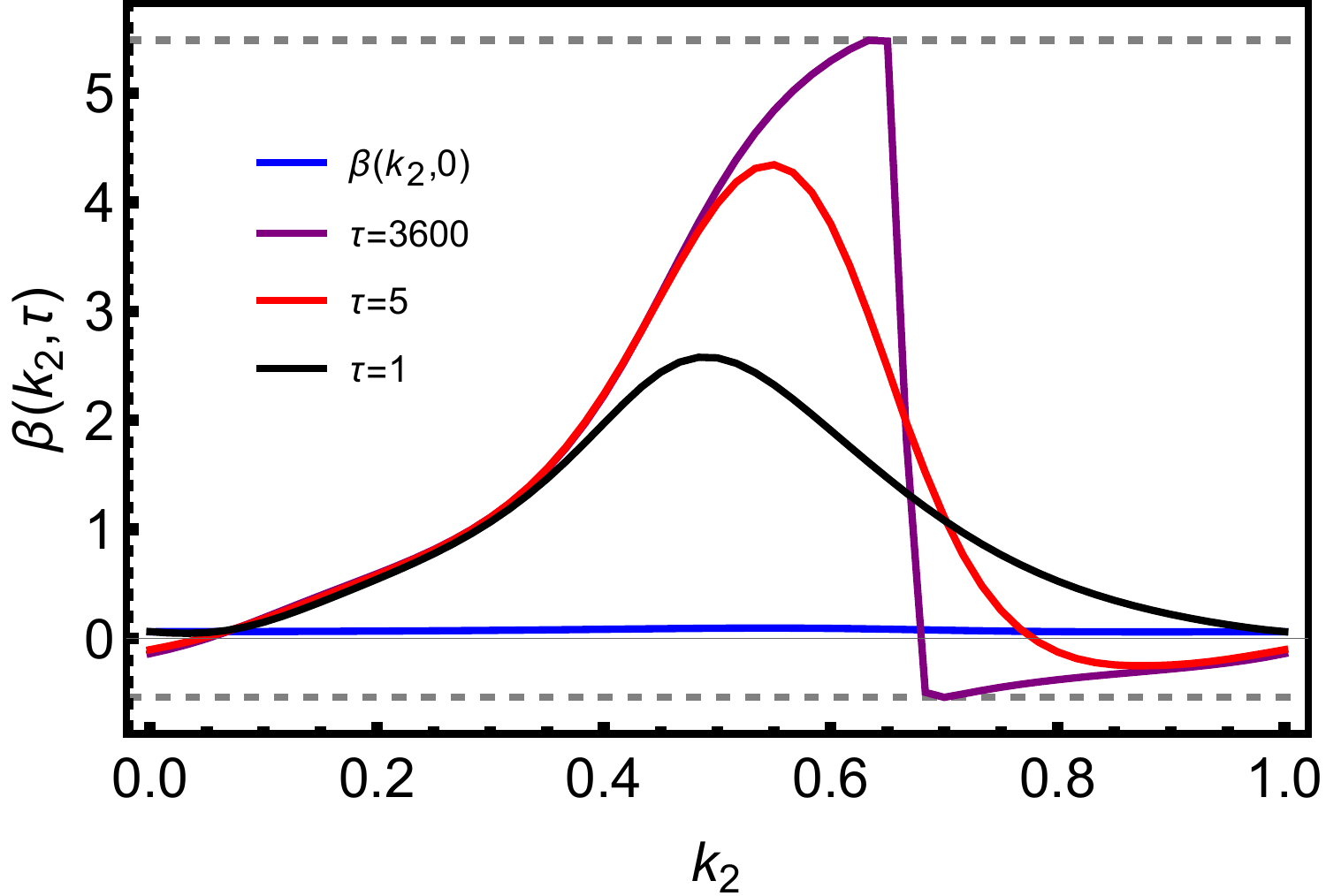}
\caption{} \label{fig_1a} 
\end{subfigure}
\begin{subfigure}{0.45\textwidth}
\centering
\includegraphics[width=1.1\columnwidth,height=0.69\columnwidth]{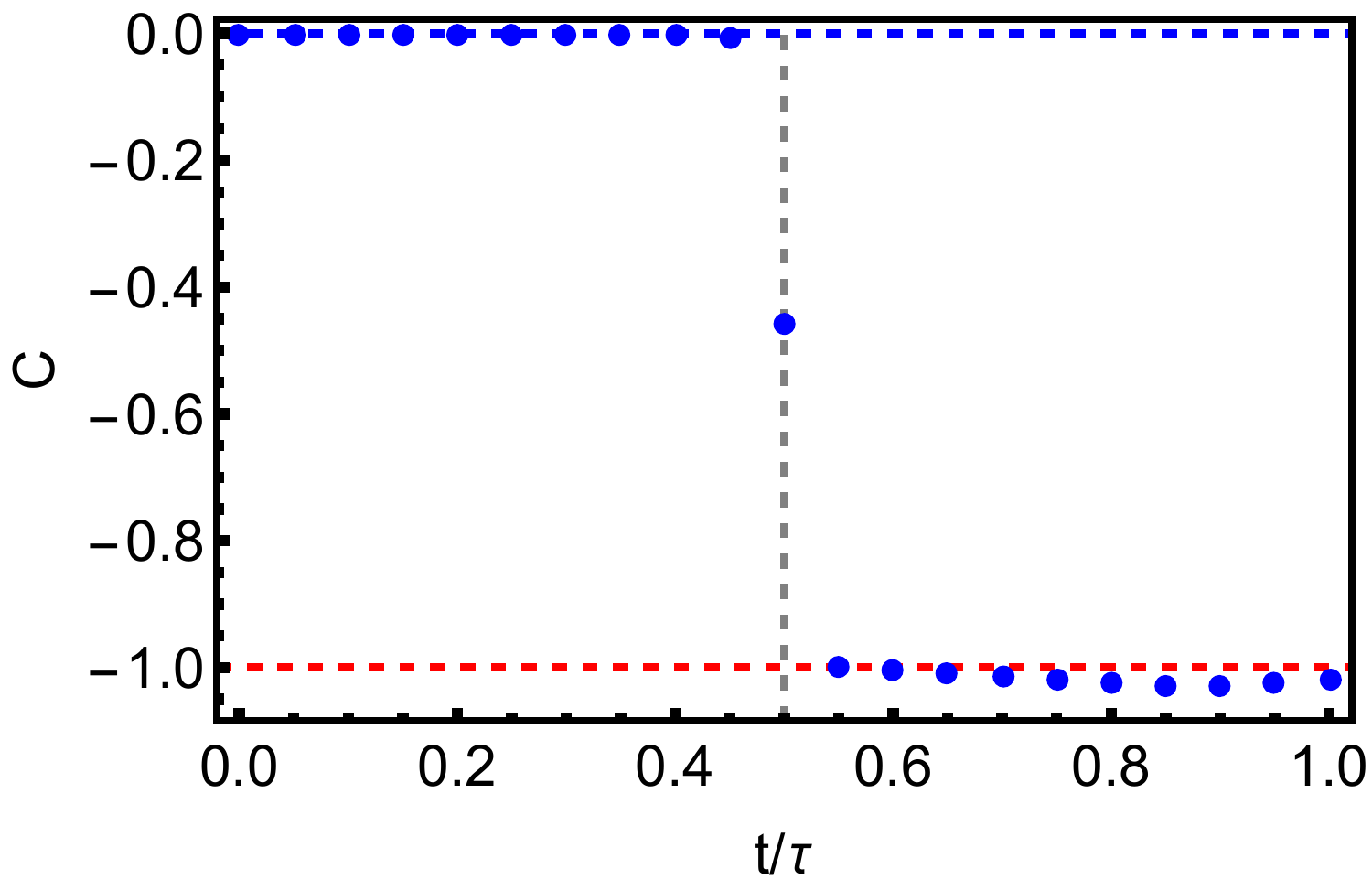}
\caption{}	\label{fig_1b}
\end{subfigure}
\cc
\caption{(a) Emergence of a sharp branch singularity in the function $\beta(k_2,\tau)$. The sharp jump in the $\beta(k_2,\tau)$ function for an adiabatic protocol (Eq.~\eqref{eq:unitary_protocol}) starting from an initial trivial state, demonstrates the topological non-triviiality of the final time evolved state. The magnitude of the jump shown by the distance between the horizontal dashed lines is $\Delta=-0.96\times2\pi$.
	The initial and the final Hamiltonian is chosen such that, nearest neighbour hopping $t_1=1.0$, next-nearest neighbour hopping $t_2=0.5$, flux through each plaquette $\phi=-\pi/2$, $M_i=3\sqrt{3}t_2+2.5$, $M_f=3\sqrt{3}t_2-2.5$.	(b) {The topological transition at $t=\tau/2$ (when the system crosses the QCP), through the topological charge of a Haldane model subjected to a linear slow quench for a $40\times 40$ lattice with, $\tau=1600$ ($\tau\simeq \tau_a$). The quench parameters are $t_1=2.0$, $t_2=1.0$, $\phi=-\pi/2$, $M_i=3\sqrt{3}t_2+2.5$, $M_f=3\sqrt{3}t_2-2.5$. (The dotted line is just a guide to eye.) }}
\end{figure*}
be included into the same equivalence class. This can be equivalently understood as since the topological invariant counts the winding of the fibre $\beta(k_2)$ over the base space $\mathcal{S}^1\equiv k_2\in[0,1]$ and merely changing the position of the topological kink does not change the homotopy class of the map. However, if a smooth gauge is not chosen and $\beta$ becomes discontinuous (and hence non-differentiable) at an inner point $k_2^*\in I$, caution must be taken while evaluating the integral in Eq.\eqref{eq:pure_chern_I},
\begin{eqnarray}\label{eq:2}
&~&\frac{1}{2\pi}\int_0^{1}\frac{d\beta(k_2)}{dk_2}dk_2= \nonumber\\
&~&\frac{1}{2\pi}\lim\limits_{\epsilon\rightarrow 0^+}\left(\int_{0}^{k_2^*-\epsilon}+\int_{k_2^*+\epsilon}^{1}\right)\frac{d\beta(k_2)}{dk_2}dk_2,
\end{eqnarray}
where we have tactically removed the isolated point $k_2^*$ where $\beta$ is not differentiable. This in a way is again equivalent to the destruction of simply-connectedness of the base manifold $I$ with respect to the map $B$, hence allowing for a non-trivial homotopy classification. By 
evaluating the integrals on the RHS of Eq.~\eqref{eq:2} we obtain,
\begin{eqnarray}
&~&\frac{1}{2\pi}\int_0^{1}\frac{d\beta(k_2)}{dk_2}dk_2 \nonumber \\
&=&\frac{1}{2\pi}\lim\limits_{\epsilon\rightarrow 0^+}\left[\beta(k_2^*-\epsilon)-\beta(k_2^*+\epsilon)\right]\nonumber\\
&=&-\Delta,
\end{eqnarray}
which is exactly the jump in $\beta(k_2)$ and therefore may be interpreted as a signature of topological non-triviality of the equilibrium system. Also, the jump $\Delta$ is a gauge invariant quantity and must be integer multiples of $2\pi$ owing to the single-valuedness of the wavefunctions at every point of $I$, i.e., 

\begin{equation}\label{eq:delta}
\Delta=2\pi\mathcal{C},~~\mathcal{C}\in\mathcal{I}.
\end{equation} 

We note, in general $\beta$ may exhibit multiple isolated discontinuities, in which case, applying a similar protocol one obtains,
\begin{equation}
\begin{split}
\mathcal{C}=\frac{1}{2\pi}\sum\limits_{\nu}\Delta_{\nu},
\end{split}
\end{equation}
where the sum is taken over all the isolated jump discontinuities of $\beta$.\\

{Moving on to}  a generic {\it  out-of-equilibrium} situation, the quantity $\mathcal{C}^U(t)$ defined in Eq. \eqref{eq:pure_chern_I} fails to be integer quantised, as a single instantaneous band may not be completely occupied far from equilibrium; {This is reflected in $\beta$ defined Eq.~\eqref{eq:pure_beta}, as a weighted average of $U(1)$ connections along a single parametric direction $k_1$, over all single-particle instantaneous bands.}
Nonetheless, for an {\it adiabatic} protocol dynamically exchanging the Chern character of two bands, the $U(1)$ connection reduces to be over the single instantaneous band which is nearly completely filled. This allows the CN to vary in time. 
Thus, in an adiabatic situation, the MEP assumes the exact $U(1)$ form,
\begin{equation}
\begin{split}
\tilde{P}_i={\rm Im}\int_{BZ}dk_{1}dk_{2}A^k_i(\ket{\phi_{ks}(t)}),
\end{split}
\end{equation} 
over the {filled} band $\alpha=s$. \\

\begin{figure*}
\begin{subfigure}{0.45\textwidth}
\includegraphics[width=\textwidth]{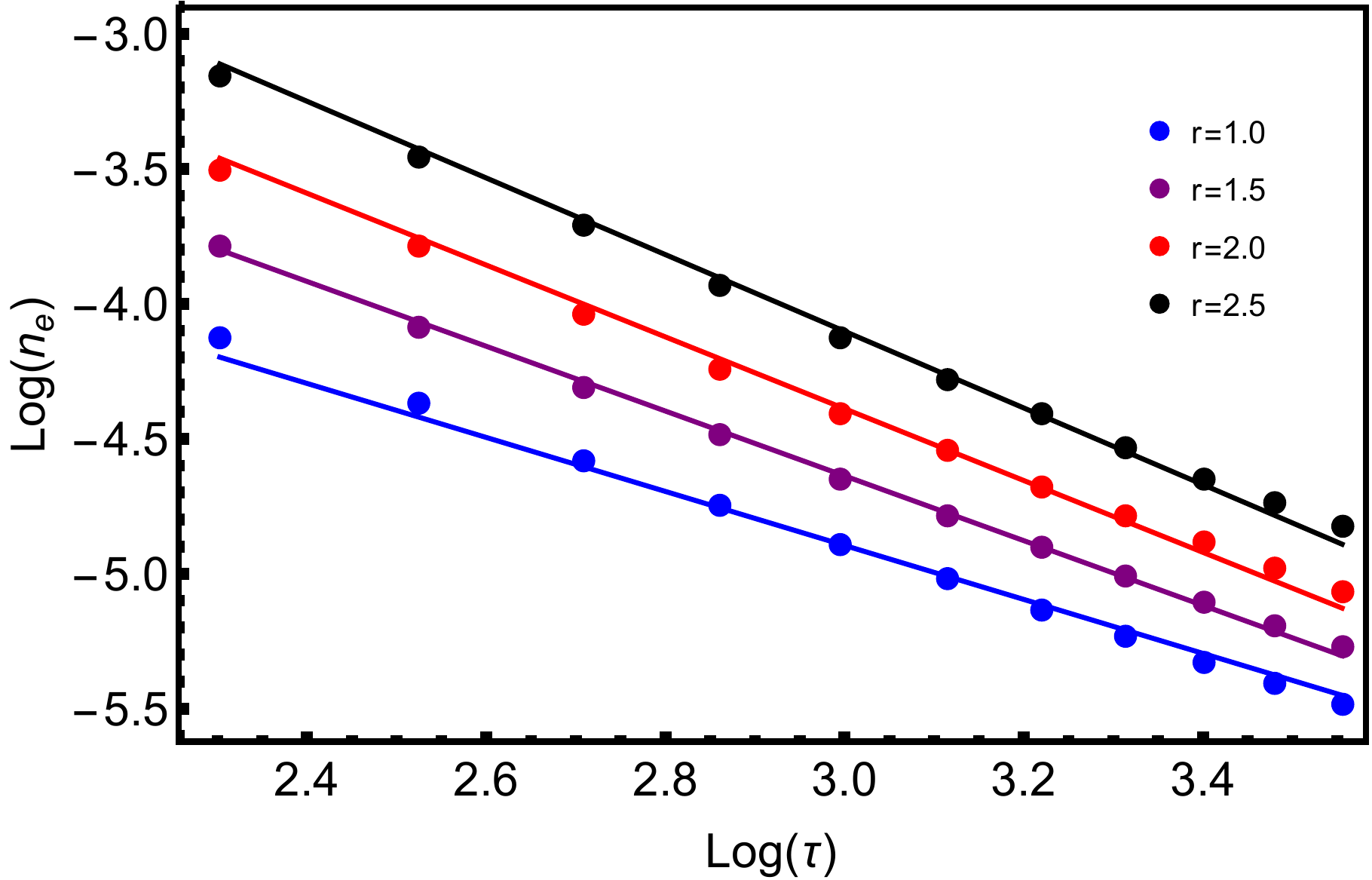}
\caption{} \label{fig_2a} 
\end{subfigure}
\quad\quad\begin{subfigure}{0.45\textwidth}
\includegraphics[width=\textwidth]{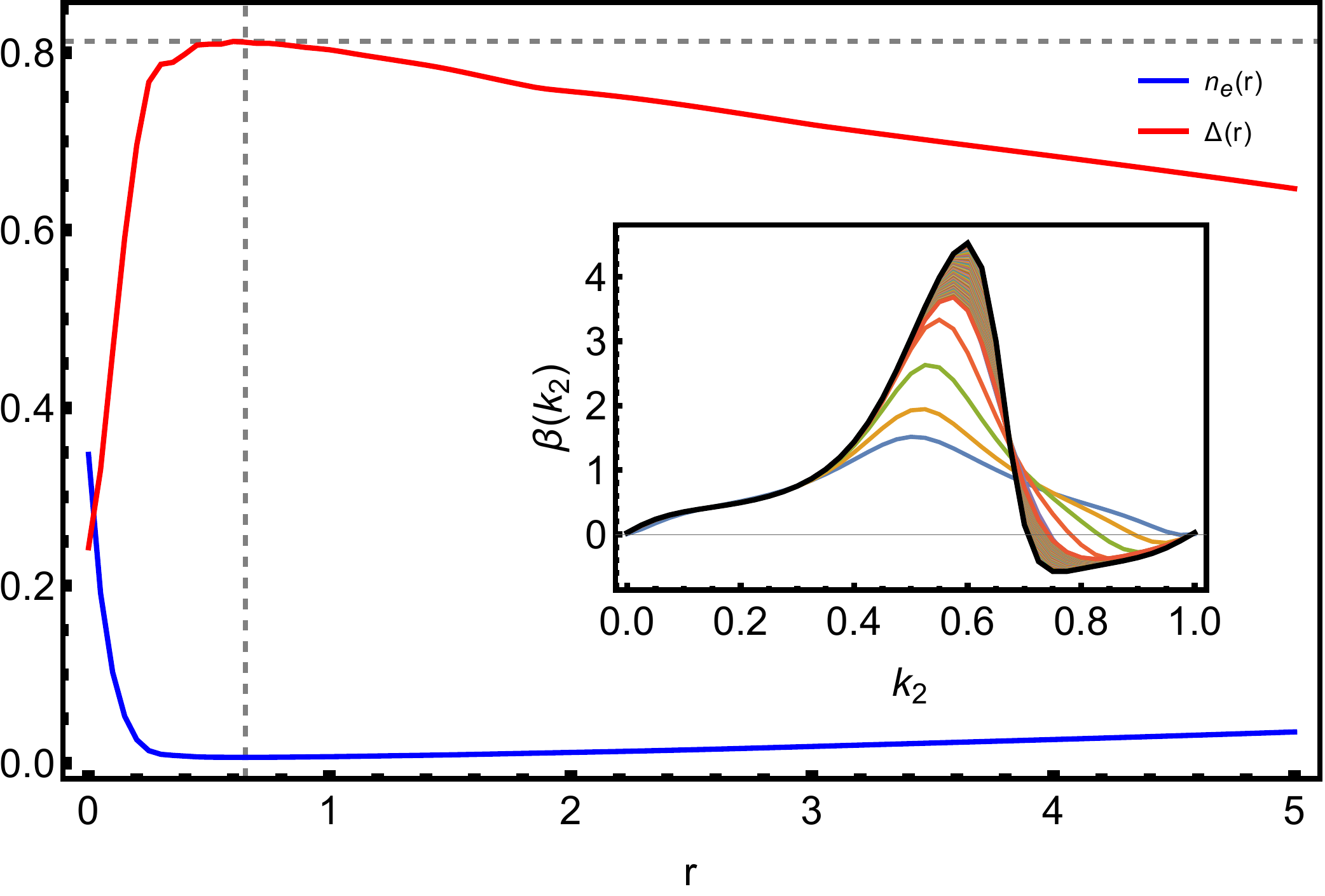}
\caption{}	\label{fig_2b}
\end{subfigure}
\caption{(Color online) (a) {The defect density (Eq.~\eqref{eq:deff:supp}) at the end of a non-linear quench (Eq.~\eqref{eq:nonl:supp}), obtained numerically
		(presented by dots), shows an excellent  agreement with the modified Kibble-Zurek scaling for different values of $r$ expressed in Eq.~\eqref{eq:kz:supp} (shown by  solid lines)}. (b) The corresponding optimal power $r_{\rm opt} \simeq 0.65$ of the protocol which minimizes the defect production at the end of the quench. (Inset) The function $\beta(k_2)$ again calculated at the end of the quench shows a considerably sharp jump of magnitude $\Delta(r)=|\Delta|_{r}/2\pi=0.81$ (Black) resembling the emergence of a topological non-triviallity ($\mathcal{C}_U\simeq -0.81$) in the final state with  $r=r_{\rm opt}$ even for a much smaller quenching time  $\tau=20.0$ for a $40\times 40$ lattice. {Notably for the linear quenching case $r=1$, one requires $\tau \sim 1600$ to
		get as close to the topological state with other quench parameters kept fixed. The parameter are}  chosen to be $t_1=1.0$, $t_2=0.5$, $M_i=3\sqrt{3}t_2+1$ (non-topological), $M_f=3\sqrt{3}t_2-1$ (topological).}
\end{figure*}

\section{Illustration with Haldane model}

To  exemplify, we consider a  {linear as well as non-linear} ramping \ct{sen08,polkovnikov08,chandran12} of  the Semenoff mass $M $ of a Haldane model in reciprocal space (see  Appendix \ref{sec_haldane} for detail), 
\begin{equation}
\begin{split}\label{eq:haldane_model}
H^k(t)=h_x(\vec{k})\sigma_x+h_y(\vec{k})\sigma_y+h_z(\vec{k},t)\sigma_z, ~~\text{with}~~ \\
\end{split}
\end{equation}

\begin{equation}\label{eq:bloch_ham}
\begin{split}
h_x(k)=-t_1\sum\limits_{i=1}^{3}\cos{\left(\vec{k}.\vec{\Delta}_{1i}\right)},\\
h_y(k)=-t_1\sum\limits_{i=1}^{3}\sin{\left(\vec{k}.\vec{\Delta}_{1i}\right)},\\
h_z(k)=M-t_2\sin{\phi}\sum\limits_{i=1}^{3}\sin{\left(\vec{k}.\vec{\Delta}_{2i}\right)},
\end{split}
\end{equation}
where $\vec{\Delta}_{1i}$ and $\vec{\Delta}_{2i}$ are the nearest neighbour and next nearest neighbour lattice vectors, $t_1$ is the nearest neighbour hopping strength, $t_2$ is the next nearest neighbour hopping strength, $M$ is the chiral symmetry breaking Semenoff mass and $\phi$ is the time-reversal breaking flux through each plaquette.

\label{sec_illustration}

\subsection{Linear quenching}

Initially ($t=0$),  the system is in a pure state  $\ket{\psi_k(0)}$ which is  the ground state of the initial (non-topological) Hamiltonian $H^k(0)$ with $M(0)=M_i$  and the final value $M(\tau)=M_f$ corresponds to a topological Hamiltonian; thus, the system is ramped across a QCP during the evolution  ({refer to Fig.~\ref{fig:phase_diagram_ap} of Appendix.~\ref{sec_haldane}}).
The protocol we propose is the following:
\begin{equation}
M(t)=M_i-(M_i-M_f)\left(\frac{t}{\tau}\right),
\label{eq:unitary_protocol}
\end{equation}
{in time $t\in[0,\tau]$.
We  {proceed to} evaluate the function $\beta(k_2,\tau)$ in the final state $\ket{\psi_k(t)}$ at $t=\tau$ generated following the evolution  under the protocol in Eq.~\eqref{eq:unitary_protocol}.
 {As shown in Fig.~\ref{fig_1a}, the function $\beta(k_2,\tau)$  develops a sharp branch singularity of nearly quantized integral multiple of $2\pi$, only when the quench approaches the adiabatic limit  {(adiabatic time scale $\tau_a\sim L^2$ for a system having {$L\times L$ sites)}; otherwise, the quantity
 ${\mathcal C}^U(t)$ loses its topological significance.}

 As discussed above, the existence of this sharp branch shift in $\beta(k_2,\tau)$, signals the topological non-triviality of the final state of the system.  
 The topological nature of the adiabatic state is also established through the emergence of a singuarity in, 
 	\begin{equation}\label{eq:field}
 	\begin{split}
 	\mathcal{F}(t)= \partial_{k_1}\tilde{A}_2(t)-\partial_{k_2}\tilde{A}_1(t),
 	\end{split}
 	\end{equation}
 	{due to an instanton at a Dirac point when the system approaches a critical point.} This arises because of the presence of a gapless point of the instantaneous Hamiltonian for large enough system sizes (see Appendix.~ 
	\ref{sec_App_MEP}). 
	{This is more precisely captured by the integral of the curvature over the complete BZ or the net flux, which acts as a closed surface integral enclosing a monopole charge $C$ (see Fig.~\ref{fig_1b}). This charge is explicitly understood as the Chern number for Chern non-trivial systems and the net flux holds a direct correspondence to Gauss Law for a $U(1)$ magnetic monopole.}}
\\


\subsection{Non-linear quench and optimal rate}

{In this section, we shall discuss} that the efficacy of adiabatic quenching protocols has been established to improve to a remarkable degree by the application of non-linear ramping schemes
\ct{sen08,polkovnikov08,chandran12} {and exploit the same for an efficient approach to the topological state}. We exemplify this advantage considering a non-linear quench in the semenoff mass of the Haldane model,
\begin{equation}\label{eq:nonl:supp}
M(t)=M_i-(M_i-M_f)\left(\frac{t}{\tau}\right)^r
\end{equation}
with an exponent $r(>0)$ from $t=0$ to $t=\tau$. The initial and final masses $M_i$ and $M_f$ are again chosen such that the initial state is trivial and the final state is topologically non-trivial as in the main text.

In the linear ramping ($r=1$) protocol, Kibble-Zurek scaling (KZS) \ct{zurek05,polkovnikov05,dziarmaga10,polokovnikov11} predicts that the defect density produced due to diabatic
excitations, e.g., in this case is the density of occupation of the excited state at the end of the quench,  defined by,
\begin{equation}\label{eq:deff:supp}
n_e=\int_{BZ}n_e^{{\bf k}}(r) d^2{\bf k},
\end{equation}
satisfies a universal scaling relation
\begin{equation}\label{eq:kzs}
n_e\sim\tau^{-\frac{\nu d}{\nu d+1}}.
\end{equation}
\begin{figure}
\centering
\includegraphics[width=1.0\columnwidth,height=0.67\columnwidth]{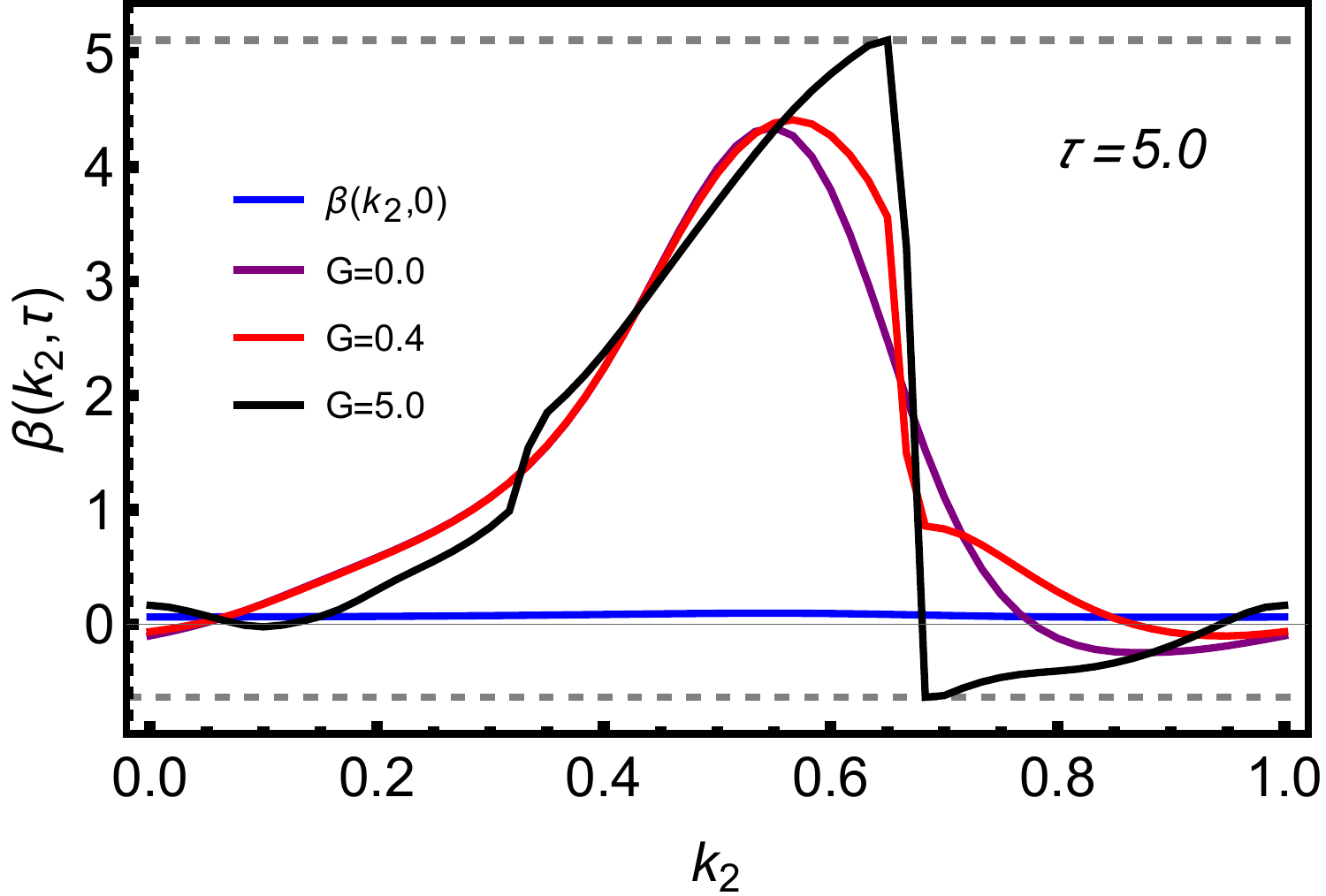}

\cc
\caption{The $\beta(k_2,\tau)$ function exhibits a sharp branch singularity in the post-quench state for a drive employing shortcut to adiabaticity (Eqs.~\eqref{eq:unitary_CD1}-\eqref{eq:unitary_CD2}) with increasing control field strength $G$. The magnitude of the jump shown by the distance between the horizontal dashed lines is $\Delta=-0.92\times2\pi$ with the set of quench parameters as in the linear quench in Fig.~\ref{fig_1a}.
	The quenching period is chosen to be $\tau=5.0$ which is much shorter than the adiabatic time scale ($\tau_a\simeq 3600$).
	Periodic boundary conditions are imposed with a grid size of $60\times 60$ lattice sites in both the figures.
}\label{fig_3} 
\end{figure}

{Here, $d$ is the spatial dimension, $\nu$ and $z$ are the correlation length and dynamical exponents associated with the quantum critical point across which the system is ramped. The defect density generated for the topological transition across a quantum critical point which the Haldane model is ramped accross is shown in Fig.~\ref{fig_2a}.}

Interestingly, for a non-linear quench, the KZS gets  modified as  the spectral minimum gap in the system gets renormalised \ct{sen08,polkovnikov08}: one arrives at
a KZS \begin{equation}
n_e(r)\sim \tau^{-\frac{\nu rd}{\nu rd+1}},
\end{equation}
for a non-linear quench of the form given in Eq.~\eqref{eq:nonl:supp}.  For the topological transition across a critical point which the Haldane model is ramped as shown in Fig.\ref{fig_2a}, the  universal critical exponents  are $\nu=1$, $z=1$ and  spatial dimension $d=2$  respectively. 
Putting these together, the defect density scales as,
\begin{equation}\label{eq:kz:supp}
n_e(r)\sim \tau^{-\frac{2r}{2r+1}},
\end{equation}
The scaling of $n_e(r)$ is verified  for the quenching protocol of Eq.~\eqref{eq:nonl:supp} in Fig.~\ref{fig_2a}. 
{Further we  highlight that there also exists an optimal power $r_{\rm opt}$ specifying the protocol such that the defect produced at the end of the quench is minimised}.
{This optimality arises because of the fact that for $r \to 0$ implying the sudden limit generates many excitations while in the  $r \to \infty$ limit  though $M(t)$ evolves very slowly close to the QCP, it changes very rapidly otherwise }\ct{polkovnikov08}. We demonstrate this precise optimization through a non-linear quench of the Semenoff mass starting from a trivial state to a non-trivial phase {and we find that the jump in $\Delta(r)$ is maximum for $r_{\rm opt}$} (see Fig.~\ref{fig_2b}).\\

We further explicitly show that remarkably, through the optimization protocol, the $\beta(k_2)$ function develops an emergent branch singularity of considerable sharpness 
even for a small quenching time $\tau$ in the optimal quench and {one finds a value of the dynamical Chern
	number ${\mathcal C}^U\simeq-1$.}

{The above numerical  observation in Fig.\ref{fig_2a}
	for a finite system can again be perfectly justified using the KZS.  Considering a non-linear quench, for a $d$-dimensional system of linear dimension $L$, it can be shown that the adiabatic limit of $\tau=\tau_{\rm a}$ scales as  $\tau_{\rm a}\sim L^{(r\nu z+1)/\nu r}$, 
	which reduces
	to $\tau_{\rm a} \sim L^{(r+1)/r}$, for the the Haldane model. This implies that for $\tau \sim \tau_{\rm a}$, the dynamics is effectively adiabatic.
	Given that attaining the adiabatic limit is essential for a perfect preparation of a topological
	state, we note that $\tau_{\rm a} \sim L^2 =1600$ for $L=40$ with the linear quench, on the contrary, for $r_{\rm opt}$ we remarkably obtain a considerable jump $\Delta(r_{opt})$ even for a small quenching time of $\tau=20$. Thus, we achieve the preparation of the  Chern topological state unitarily with a high fidelity at a much smaller value of  $\tau$ for the optimal ramping protocol. }
{With increasing system size, the value $r_{\rm opt}$ would change and consequently the jump in $\beta(k_2)$ would be even sharper.}

\subsection{Counter-diabatic protocol}
During the passage through a gapless QCP, the adiabaticity criteria necessarily breaks down in the thermodynamic limit and diabatic
excitations are inevitable. Nevertheless, the application of a control field \cite{anatoli19,anatoli17,marin14} may  
allow one to approach adiabaticity quicker {even for large systems, than within the protocol in Eq.~\ref{eq:unitary_protocol}}, thereby, allowing for  a much more efficient  preparation of a topological state even for $\tau\ll\tau_a$.\\

\begin{figure*}
\begin{subfigure}{0.45\textwidth}
\includegraphics[width=\textwidth,height=0.2\textheight]{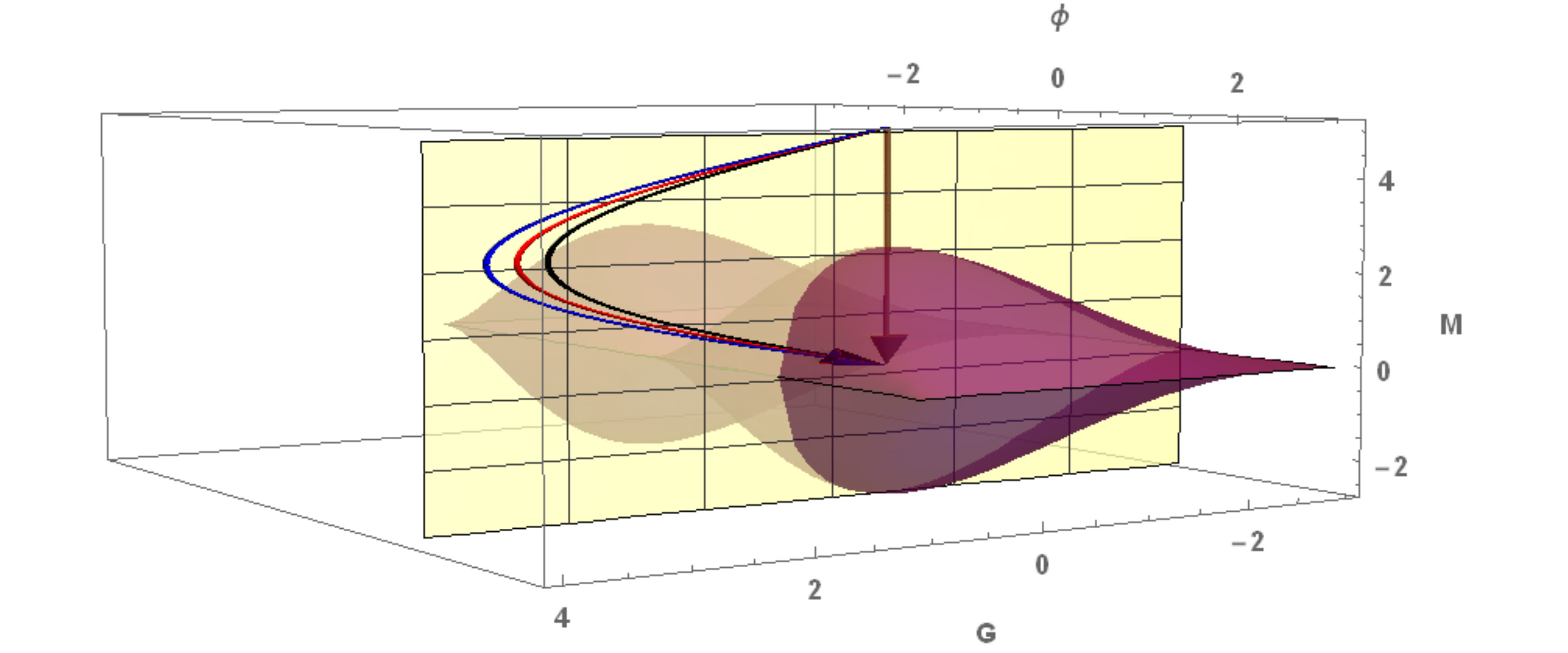}
\caption{.} \label{fig_4a} 
\end{subfigure}
\quad\quad\begin{subfigure}{0.45\textwidth}
\includegraphics[width=\textwidth]{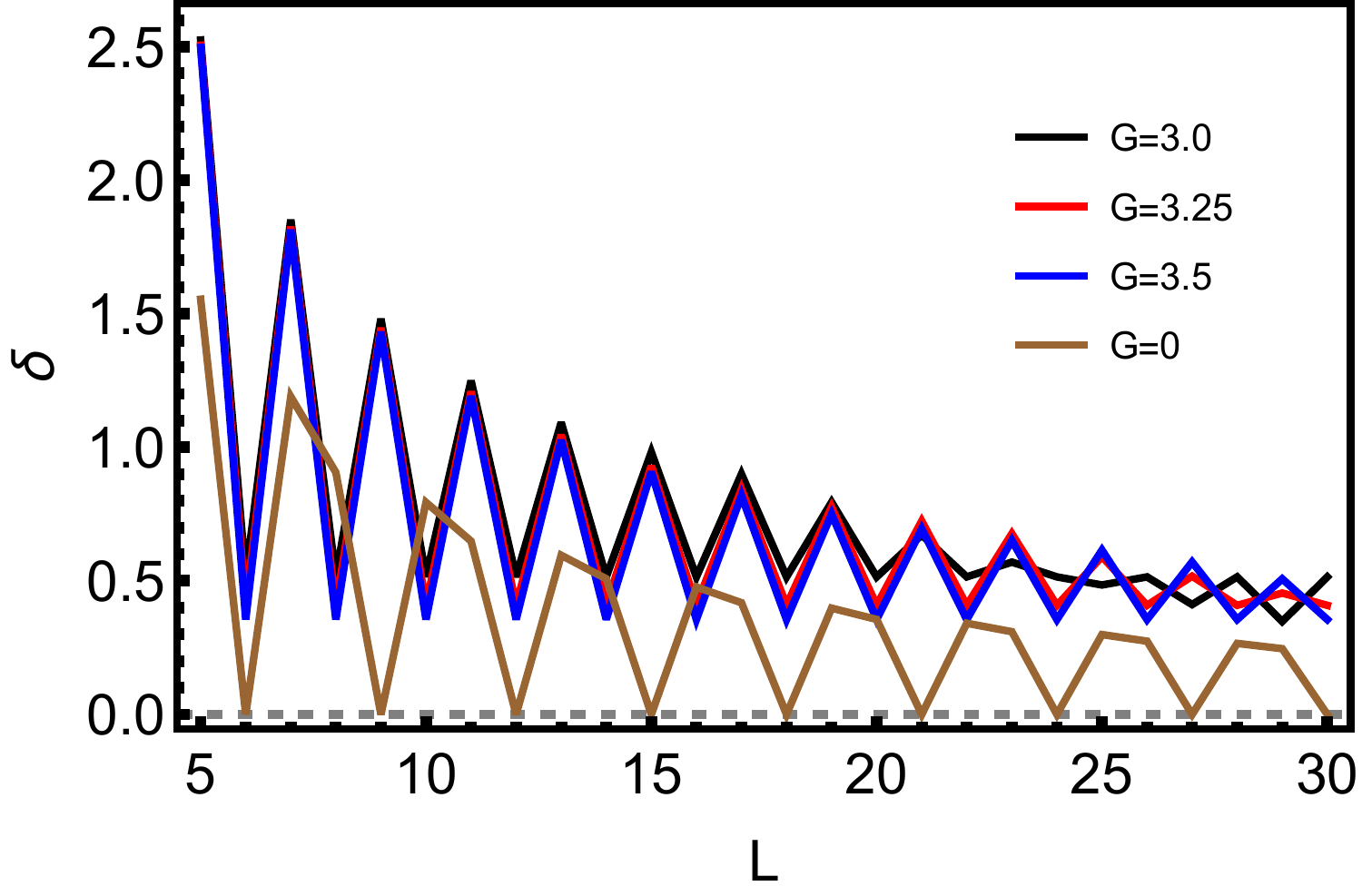}
\caption{}	\label{fig_4b}
\end{subfigure}
\caption{(a) The 2D critical surface (purple) in a continuum system parametrised by $M$, $\phi$ and the counter-diabatic field $G$ and the trajectory of different quenches (1D lines) for different values of $G$. (b) The minimum gap in system vs system size $L\times L$ at times when the quench trajectory intersects the critical surface for different values of  G.  The minimum gap encountered during the topological transition is enhanced for a non-zero $G$ when compared to the protocol for $r=1$, where $G=0$, for  the same system size.}
\end{figure*}
The protocol we propose is the following:
\begin{eqnarray}
	\label{eq:unitary_CD1}
		H^k(t)&=&h_x(\vec{k})\sigma_x+h_y(\vec{k})\sigma_y+h_z(\vec{k},t)\sigma_z-B_x(t)\sigma_x, \nonumber\\
		M(t)&=&M_i-(M_i-M_f)\frac{t}{\tau},
		\end{eqnarray}
where the control (counter-diabatic) field is chosen as:
\begin{equation}\label{eq:unitary_CD2}
B_x(t)=G\sin{\left(\frac{\pi t}{\tau}\right)},t\in[0,\tau]; ~B_x(0)=B_x(\tau)=0. 
\end{equation}
{ The CD mass in Eq.~\eqref{eq:unitary_CD1} is generated numerically in the real space lattice by inducing an anisotropy  in the nearest neighbour hoping strength
 as we have discussed in Appendix \ref{App_CD}.}\\

Under protocol \eqref{eq:unitary_CD1}, again the initial system is in a trivial state while the final is expected to be topological one.
Starting from the  ground state of the initial Hamiltonian, 
we probe the emergence of topology in the out of equilibrium state. In Fig.~\ref{fig_3}, we observe that once again the post-quench state develops a sharp branch singularity showing near quantisation of the jump $\Delta$ (i.e., $C^U(\tau)\simeq -1$) however in a much shorter duration of quench than that in the case of a linear ramp. { The advantage of the CD term is that it expands the phase diagram into an additional parametric direction. Importantly,  the minimum gap encountered during the topological transition is enhanced for a non-zero $G$ when} compared to the protocol Eq.~\eqref{eq:unitary_protocol} (where $G=0$) for same system size (see also  for 
	illustration). This allows one to maintain adiabaticity for shorter quench times in considerably larger systems. {Thus, even though adiabiticiy  necessarily breaks down
	in crossing the QCP in a thermodynamically large system, the CD protocol provides an efficient method for experimentally  relevant finite systems.} \\

To particularly understand the role of the CD driving, it is essential to note that it allows adiabaticity in considerably large system sizes for small quenching times that are not feasible in simple annealing protocols. However, the quench is bound to cross a critical point in a thermodynamically large continuum system if one intends to change the topology of the system. In Fig.~\ref{fig_4a}, we show the  critical surface of the system as a function of all the free parameters and show that the counter-diabatic (CD) quench trajectory necessarily crosses a critical point for different values of the CD field $G$. However, as shown in Fig.~\ref{fig_4b}, the minimum gap $\delta$ in the system during the transition point is considerably higher in the CD quench as compared to a simple linear quench. This clearly illustrates the advantage of using a CD protocol to suppress excitations even in system sizes hosting a very small gap at the transition point. \\


\subsection{Bulk-boundary correspondence}

\label{sec_bulk_boundary}
 The {measurable} identity of the topological nature of the post quench state  is manifested in the emergence of localized edge currents $J_L^x$ under semi-periodic boundary conditions in a system having $L\times L$ atoms, as demonstrated in Fig.~\ref{fig_5}.
 To evaluate the edge currents we impose semi-periodic boundary conditions on the 2D lattice. Generically, as defined above, the single particle current can be decomposed as (see
 also Appendix \ref{App_current}),
 \begin{equation}\label{curr}
 \braket{\vec{J}_{SS}}=\braket{\vec{J}_N}+\braket{\vec{J}_{NN}},
 \end{equation}
 where $\vec{J}_{N}$ and $\vec{J}_{NN}$ are the nearest neighbour and the next nearest neighbour current operators respectively,
 \begin{equation}
 \begin{split}
 \braket{J_{Nn}^x}=\sum\limits_{m}t_1\braket{a_{n}^{\dagger}a_m}-hc \\
 \braket{J_{NNn}^x}=\sum\limits_{m}t_2\braket{a_{n}^{\dagger}a_m}-hc,
 \end{split}
 \end{equation}
 where $\braket{J_{Nn}^x}$($\braket{J_{NNn}^x}$) is the nearest(next nearest) current at the $n^{th}$ site and the summation indices extends over all nearest (nest-nearest) neighbour sites to the $n^{th}$ site. Considering the lattice to be periodically wrapped in the x-direction (see Appendix \ref{App_edge}) while being open in the y-direction, one obtains two arm-chair edges at the ends of the cylinder. We compute the total horizontal current flowing in the periodic x-direction on one of the arm chair edges $J^x_L$ for a $L\times L$ lattice.
 The  { existence of the localised edge current} therefore bears the signature of the { post-quench bulk-boundary correspondence} both under adiabatic and CD dynamics. {In Appendix \ref{App_edge}
 , we show that the current is indeed
 localised at the edges and decays rapidly in the bulk. Further, the transition reflected in the edge-behavior is expected to be sharper with increasing system size.}
  \\	

\section{Outlook and Concluding comments}

\label{sec_outlook}
In conclusion, we have achieved  the dynamical preparation of topological states 
 of a CI within a unitary set up.
%
 The dynamical CN evolves with time unlike that defined in Ref. \ct{rigol15} and  assumes an  integer-quantised value, though
 not for an arbitrary protocol,  nevertheless
for a perfectly  adiabatic evolution.

{On a comparative note, crossing a {gapless} QCP is inevitable in a topological quench both for linear  and non-linear ramping protocols in the thermodynamic limit. However, we establish an improvement in the branch singularity in the final topological state through an optimal non-linear ramp for a much shorter quench time as compared to the adiabatic time-scale in a large but finite size system. Interestingly though, we establish a remarkable improvement in the branch singularity in the final topological state through a CD protocol for a much shorter quench time as compared to the adiabatic time-scale. }


{ Further, the CD protocol we propose, was not reported before, to the best of our knowledge and at the same time can be experimentally generated in graphene and borophene lattices by applying anisotropic strains in particular bond directions or through dynamical gap manipulations as explored theoretically and experimentally in \cite{neto09,peeters13,olivia16,olivia13,hua11,levy10,naumis19,peeters16}. Furthermore, we reiterate the experimental possibility of directly measuring the many-body Chern number has already been proposed recently through the measurements of correlations \ct{cian20}.} Interestingly, as the MEP can be written in terms of single particle correlators, it would be interesting to probe its long time thermalization properties in fast quenches. One may also proceed to study the many-body topological nature of generelised Gibbs state in quenched integrable systems. Although, herein we basically introduce the many-body invariant through the MEP and deal with the adiabatic scenario, future studies are neccessary to comprehend the scope of the many body Chern invariant.

 \begin{figure}
 \centering
 
 \includegraphics[width=\columnwidth,height=0.65\columnwidth]{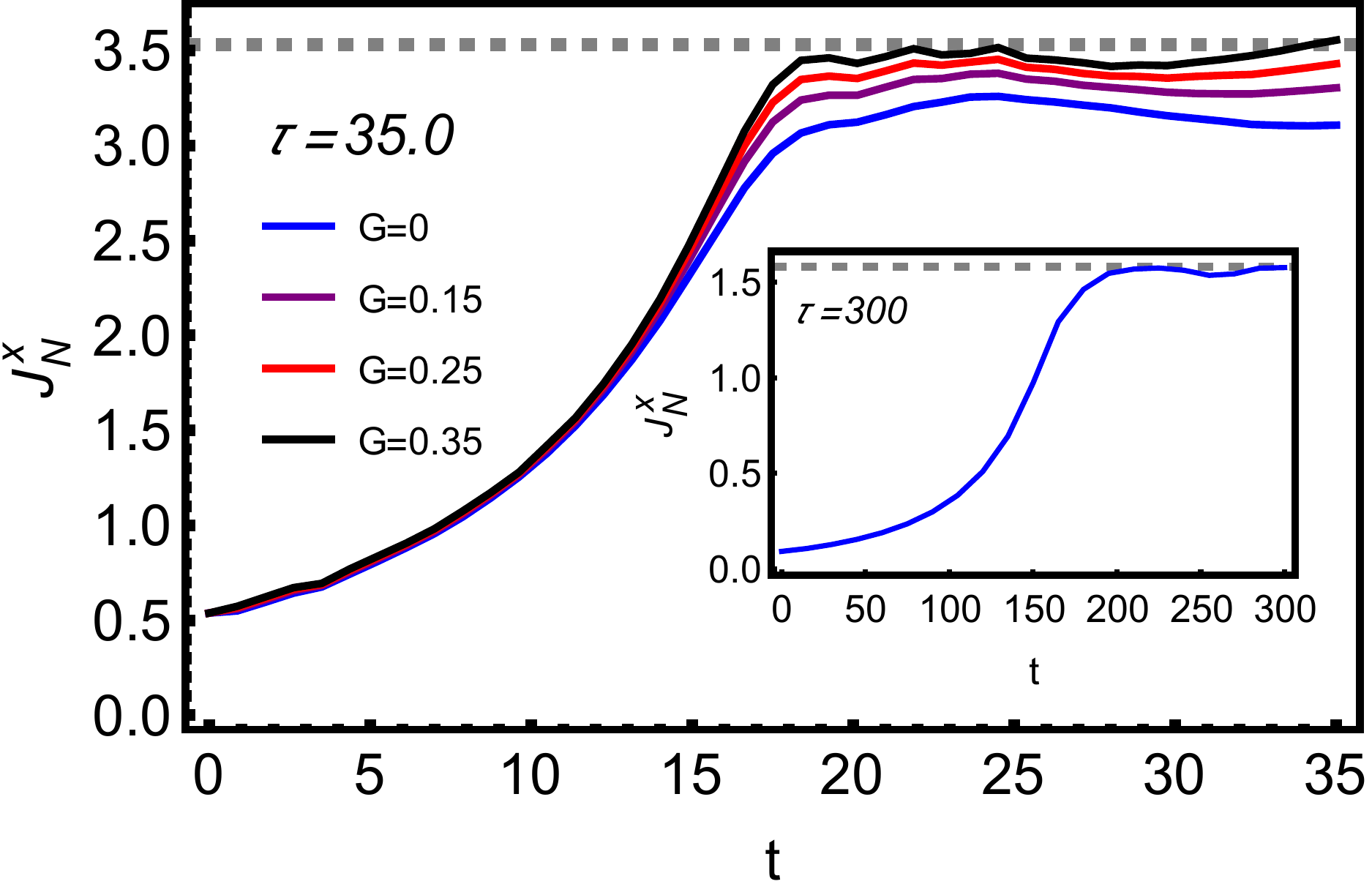}
\cc
 \caption{The time evolution of the magnitude of the edge-current $J_N^x$ through an arm-chair edge of the Haldane model with semi-open boundary conditions (periodic about the x-direction and open in y-direction) under a linear quench with the counter diabatic mass generation as in Eq.~\eqref{eq:unitary_CD1} with a $18\times 18$ lattice.
 	The quench parameters are $t_1=2.0$, $t_2=1.0$, $\phi=-\pi/4$, $M_i=3\sqrt{3}t_2+2.5$, $M_f=0$, for a $18\times 18$ lattice. 
 	$G$ is the strength of the anisotropic hopping (see Appendix.~\ref{App_CD}) introduced in the real space lattice which in turn generates a CD mass.
 	(Inset) The adiabatic evolution of the edge-current vide the protocol in Eq.~\eqref{eq:unitary_protocol} with parameters same as in Fig.~\ref{fig_2b} for a $16\times 16$ lattice. At $t=\tau$,  the edge-current (solid curve) thermalizes to its equilibrium topological value (dashed line) in both protocols.}\label{fig_5}
 \end{figure}

\begin{acknowledgments}
	We specially acknowledge  Utso Bhattacharya, Arijit Kundu, Somnath Maity, Sougato Mardanya, Anatoli Polkovnikov and Diptiman Sen for helpful discussions and critical comments. We thank Sourav Bhattacharjee for critical reading of the manuscript. SB acknowledges PMRF, MHRD India for financial assistance. AD acknowledges financial support from SPARC program, MHRD, India. This research was supported in part by the International Centre for Theoretical Sciences (ICTS) during a visit for participating in the program -  Thermalization, Many body localization and Hydrodynamics (Code: ICTS/hydrodynamics2019/11).\\

\end{acknowledgments}

\appendix

\section{Macroscopic polarisation, Many-Body Chern number and its out-of-equilibrium generalisation}

\label{sec_App_MEP}

We evaluate the macroscopic electric polarisation vector \ct{resta98,vanderbilt09} of the system in the directions of the lattice basis vectors (see Fig.~\ref{fig_1a_ap}),
\begin{equation}
\vec{P}=\sum\limits_{i}P_{\hat{i}}{\bf \hat{a}_i},
\end{equation}
where $P_{\hat{i}}=\left<\hat{X}_i\right>$, and ${\bf \hat{a}_i}$ are the lattice basis vectors. The quantity  $\hat{X}_i=\sum\limits_{n}x_{i}^n\hat{a}^{\dagger}_n\hat{a}_n$ is the many-body position operator where $x_i^n$ denotes the coordinate of an atom at the $n^{th}$ site along the $i^{th}$ lattice direction with $a_n^{\dagger}$ being the corresponding  fermionic creation operator at that site. The expectation is taken over a fermionic many body state. The  {momentum} translation operator in the $i^{th}$ direction under periodic boundary conditions,
\begin{equation}
\hat{T_{i}}(\delta_i)=e^{i\delta_{i}\hat{X}_{i}},
\end{equation}
where we choose $\delta_i=2\pi/L_i$, $L_i$ being the dimension of the system in the $i^{th}$ direction. The periodicity of the exponential enforces periodic boundary conditions on the lattice. Therefore, under periodic boundary conditions and in the thermodynamic limit, the macroscopic polarisation of the system assumes the following form,
\begin{equation}\label{Seq:pol_def_ap}
P_i={\rm Im}\ln\left<\hat{T}_i\right>,
\end{equation}
where the expectation is taken over the full many-body state of the system.
{In the thermodynamic limit, this compactified definition of the macroscopic polarisation reduces to the conventional bulk polarisation of the system. This is evident from the fact that, for a many-particle pure state, $\ket{\Psi}$ (which is a slater determinant of the occupied single-particle states),
	\begin{equation}
	P_i={\rm Im}\ln\left<\hat{T}_i\right>={\rm Im}\ln \det{U}={\rm Im}\ln e^{{\rm Tr}\ln U},
	\end{equation}
	where the matrix $U$ contains all the overlap of the single-particle matrix $T_i$ between the occupied single particle states, i.e.,
	\begin{widetext}
	\begin{equation}
	U_{mn}=\bra{\psi_m}T_i\ket{\psi_n}\implies (U)_{k\alpha,k^{\prime}\alpha}=\langle{\psi_{k_i+\delta_i,\alpha}|\psi_{k,\alpha}}\rangle98	\simeq e^{-i\left(A_{i}^k\right)_{\alpha\alpha}\delta_i},
	\end{equation}
	\end{widetext}
	where $k$ denotes the single-particle momenta while $\alpha$ is the single-particle band indices and $\left(A_{i}^k\right)_{\alpha\alpha}$ is the $U(1)$ connection of the $\alpha^{th}$ occupied band. 
	In the thermodynamic limit ($\delta_i\rightarrow 0$), retaining only terms of linear order in $\delta_i$, one obtains,
	\begin{equation}\label{Seq:pure_pol_1_ap}
	P_i=\sum\limits_{\alpha}{\rm Im}\int_{BZ}\langle{\psi_{k,\alpha}|\partial_{k_i}|\psi_{k,\alpha}}\rangle dk_{1}dk_{2},
	\end{equation}
	which is the bulk macroscopic polarisation of the system.\\
	
	The Chern invariant conventionally defined as,
	\begin{equation}
	C=\frac{1}{4\pi}\int_{BZ}dk_1dk_2\left[\partial_{k_1}\langle{\psi_k|\partial_{k_2}|\psi_k}\rangle-\partial_{k_2}\langle{\psi_k|\partial_{k_1}|\psi_k\rangle}\right],
	\end{equation}
	can be recast to the form,
	\begin{widetext}
	\begin{equation}
	C=\frac{1}{2\pi}\int_{k_{20}}^{k_{20}+1}dk_2\partial_{k_2}\int_{k_{10}}^{k_{10}+1} dk_1 \langle{\psi_k|\partial_{k_1}|\psi_k}\rangle=-\frac{1}{2\pi}\int_{k_{20}}^{k_{20}+1}dk_2\partial_{k_2}\beta(k_2),
	\end{equation}
	\end{widetext}
	where,
	\begin{equation}
	\beta(k_2)=-{\rm Im}\int_{k_{10}}^{k_{10}+1} dk_1 \langle{\psi_k|\partial_{k_1}|\psi_k}\rangle.
	\end{equation}
	The Chern number therefore essentially counts the $U(1)$ winding of the map,
	\begin{equation}
	\mathcal{S}^1:k_2\in[0,1]\rightarrow\mathcal{S}^1:\beta(k_2).
	\end{equation}

	In the main text, we consider   an arbitrary unitary   drive starting from an initial eigenstate state $\ket{\psi(0)}$ of the Chern insulator (this ensures half-filling of the initial single-particle states)
	in the non-topological phase with ${\mathcal C}=0$ (as shown in Fig. \ref{fig:phase_diagram_ap}) such that the time evolved state is,
	\begin{equation}
	\ket{\psi(t)}=U(t,0)\ket{\psi(0)},
	\end{equation}
	where $U(t,0)$ is the evolution operator generated by an instantaneous hermitian Hamiltonian $H(t)$.  
	Translating to  Fourier space, the instantaneous eigenmodes $\ket{\phi_{k\alpha}(t)}$ of the instantaneous Hamiltonian $H_k(t)$ 
	satisfy,
	\begin{equation}
	H_k(t)\ket{\phi_{k\alpha}(t)}=E_{k\alpha}(t)\ket{\phi_{k\alpha}(t)},
	\end{equation}
	with eigenvalues $E_{k\alpha}(t)$, for all $k\in BZ$. 
	and  $\alpha$ denotes the band index.

	As discussed in Eq.~\eqref{Seq:pure_pol_1_ap}, the electric polarisation in the $i^{th}$ direction for an arbitrary pure quantum many-body state $\ket{\chi}$ reduces to the average of the quantity,
	\begin{equation}\label{eq:pure_con_ap}
	\Lambda_i^k=\sum\limits_{\alpha}A^k_i(\ket{\chi_{k\alpha}}),
	\end{equation}
	over the complete Brillouin zone and summed over all the occupied single particle states $\ket{\chi_{k\alpha}}$. Here, $A^k_i(\ket{\chi_{k\alpha}})$ is the $U(1)$ gauge connection on the state $\ket{\chi_{k\alpha}}$ i.e.,
	\begin{equation}
	A^k_i(\ket{\chi_{k\alpha}})=\braket{\chi_{k\alpha}|\partial_{k_i}|\chi_{k\alpha}}.
	\end{equation}

	In the out-of-equilibrium situation, we extend the quantity defined in Eq.~\eqref{Seq:pure_pol_1_ap} 
	as a weighted average over the instantaneous single-particle bands,
	\begin{equation}\label{eq:ct_def:supp_ap}
	\tilde{P}_i=\sum\limits_{\alpha}{\rm Im} \int_{BZ}dk_{1}dk_{2}n_{\alpha}^k(t)A^k_i(\ket{\phi_{k\alpha}(t)}),
	\end{equation}
	where the weights $n_{\alpha}^k(t)$ are the time dependent population of the $\alpha^{th}$ instantaneous band i.e.,
	\begin{equation}
	n^k_{\alpha}(t)=\braket{\psi_k(t)|c^{\dagger}_{k\alpha}(t)c_{k\alpha}(t)|\psi_{k}(t)},
	\end{equation}
	ere $c_{k\alpha}(t)$ and $c_{k\alpha}^{\dagger}(t)$ are the annihilation
	and creation operators respectively, of the eigenmodes of the
	instantaneous Hamiltonian $H_k(t)$, i.e., 
	$c_{k\alpha}^{\dagger}(t)\ket{0}=\ket{\phi_{k\alpha}(t)}$,
	where $\ket{0}$ is fermionic vacuum.
	$\tilde{P}_i$ is the weighted average of the electric polarisation in each band of the time-evolved Hamiltonian $H(t)$; the weights being precisely the time dependent population of each band.  { We re-iterate that the topological invariant perfectly reduces to the convntional Chern number in an equilibrium setting under half-filing.}\\

\begin{figure*}
\begin{subfigure}{0.45\textwidth}
\includegraphics[width=\textwidth]{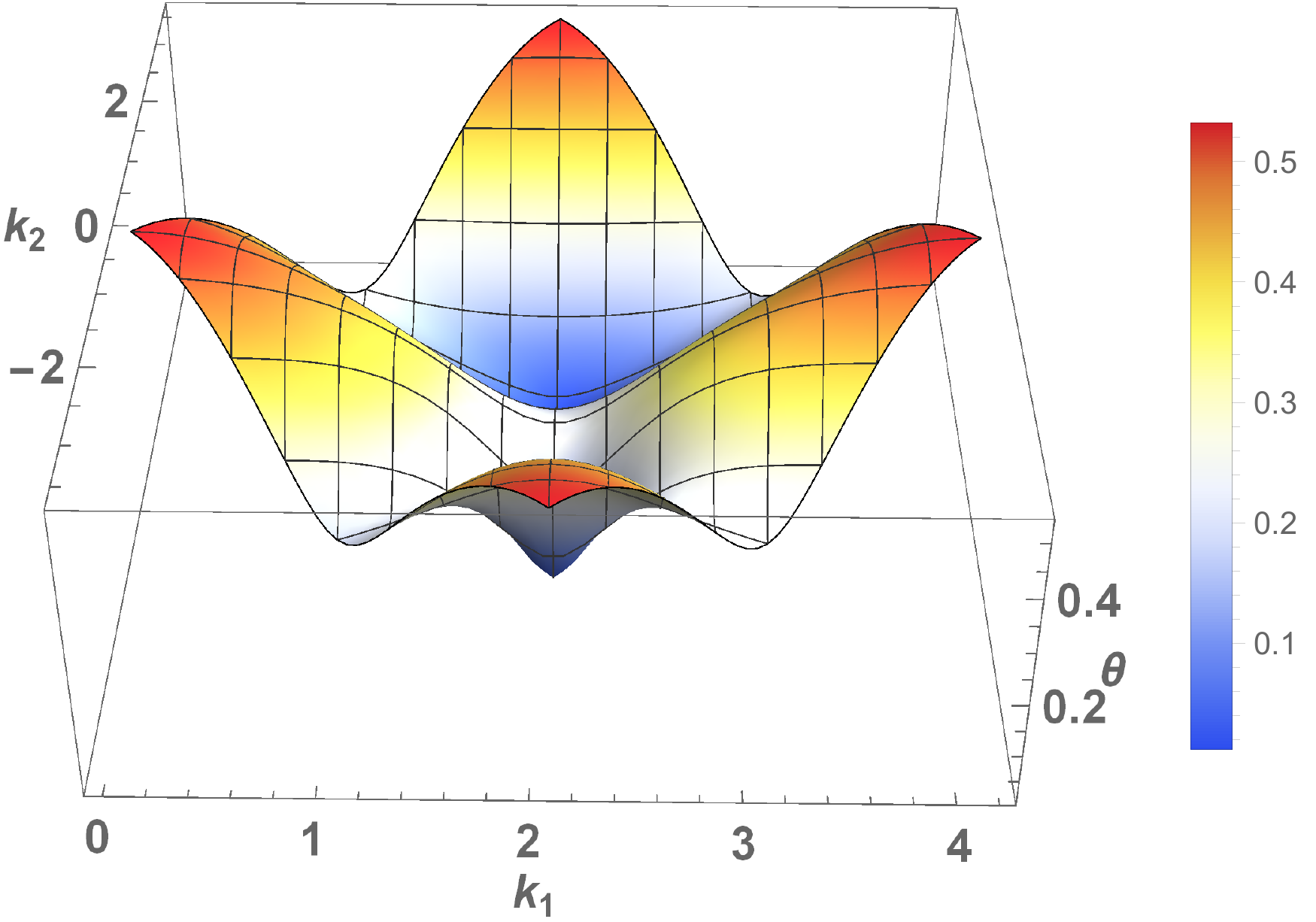}
\caption{} \label{fig_1a_res} 
\end{subfigure}
\quad\quad\begin{subfigure}{0.45\textwidth}
\includegraphics[width=\textwidth]{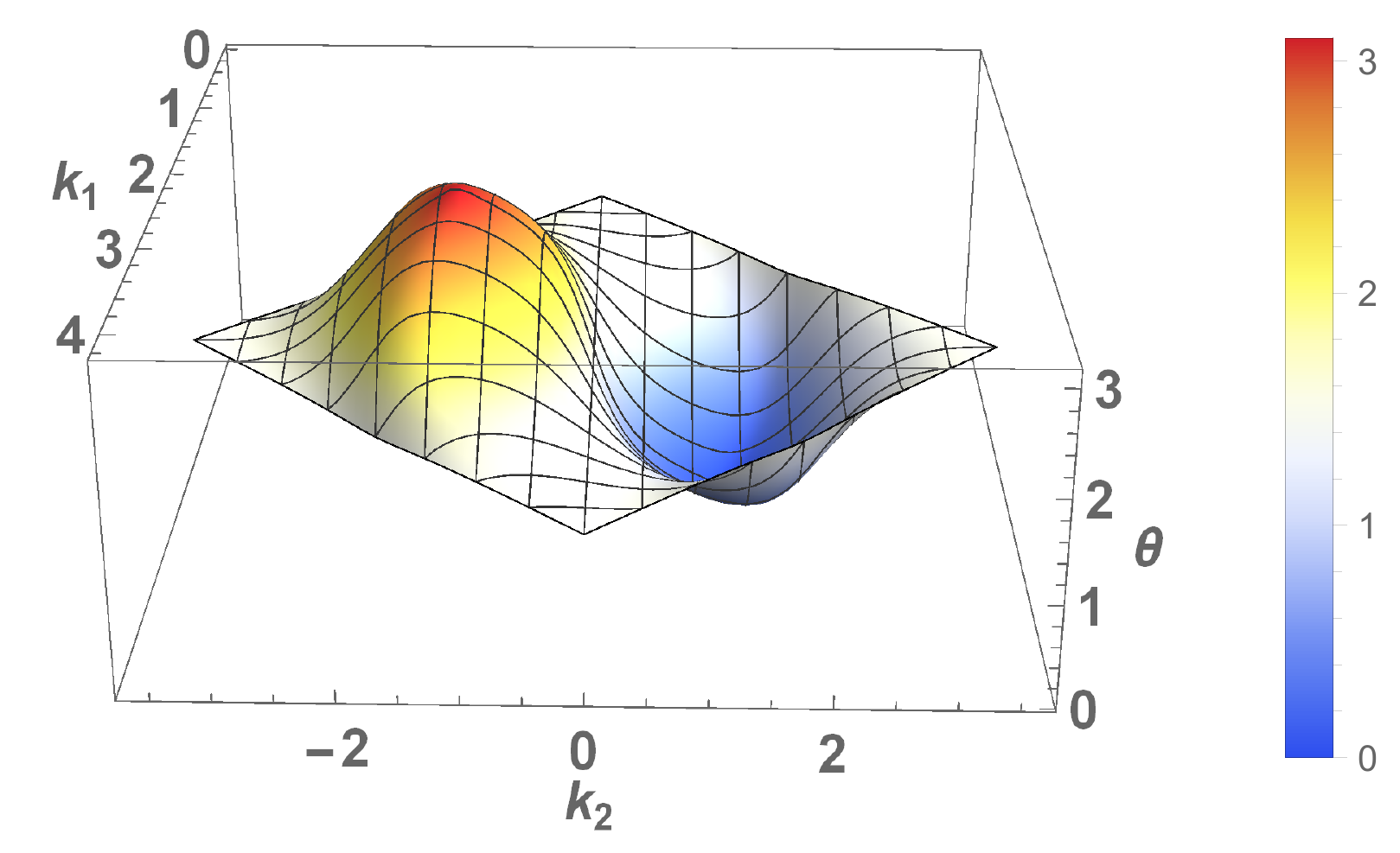}
\caption{}	\label{fig_1b_res}
\end{subfigure}
\quad\quad\begin{subfigure}{0.45\textwidth}
\includegraphics[width=\textwidth]{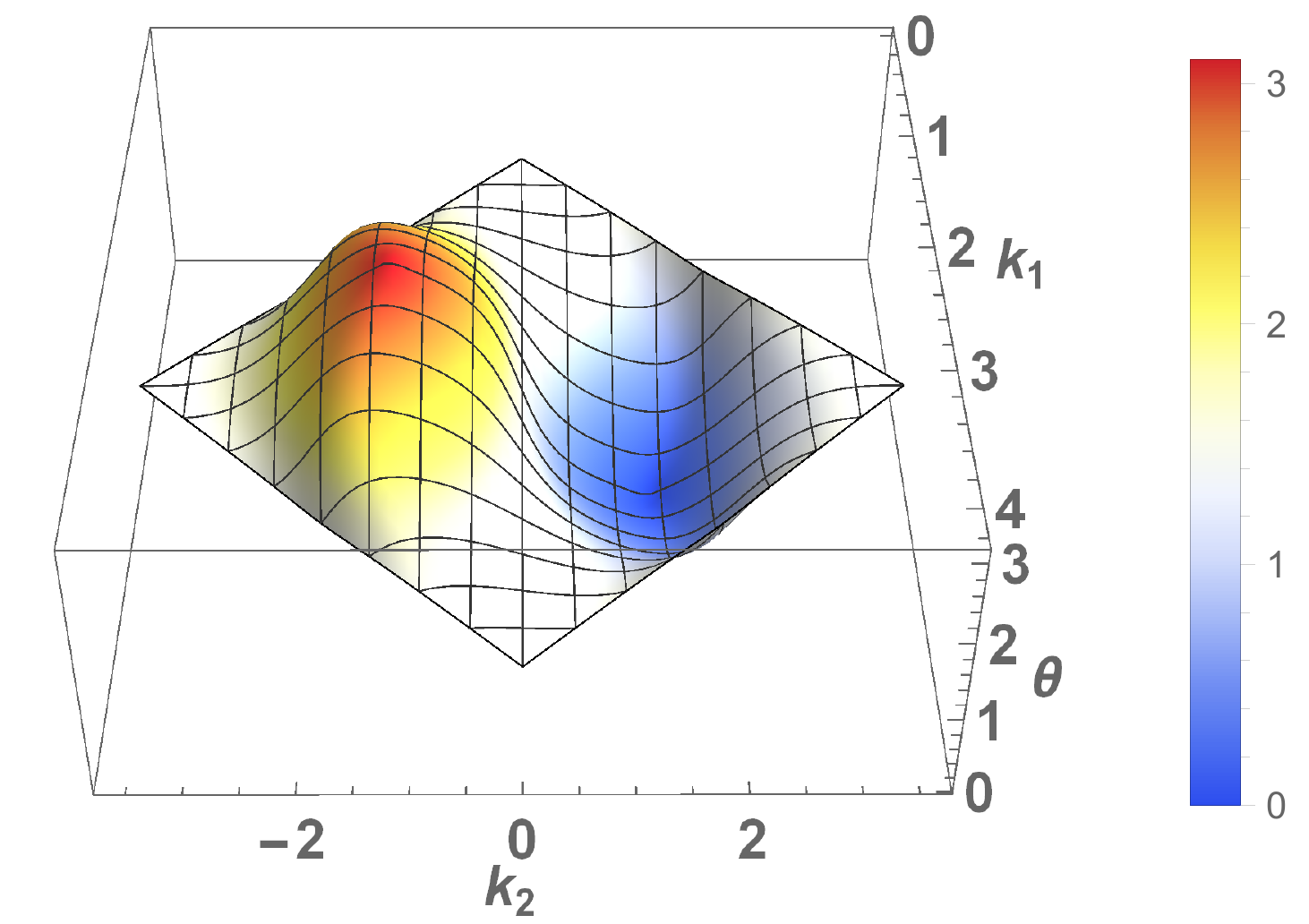}
\caption{}	\label{fig_1c_res}
\end{subfigure}
\quad\quad\begin{subfigure}{0.45\textwidth}
\includegraphics[width=\textwidth]{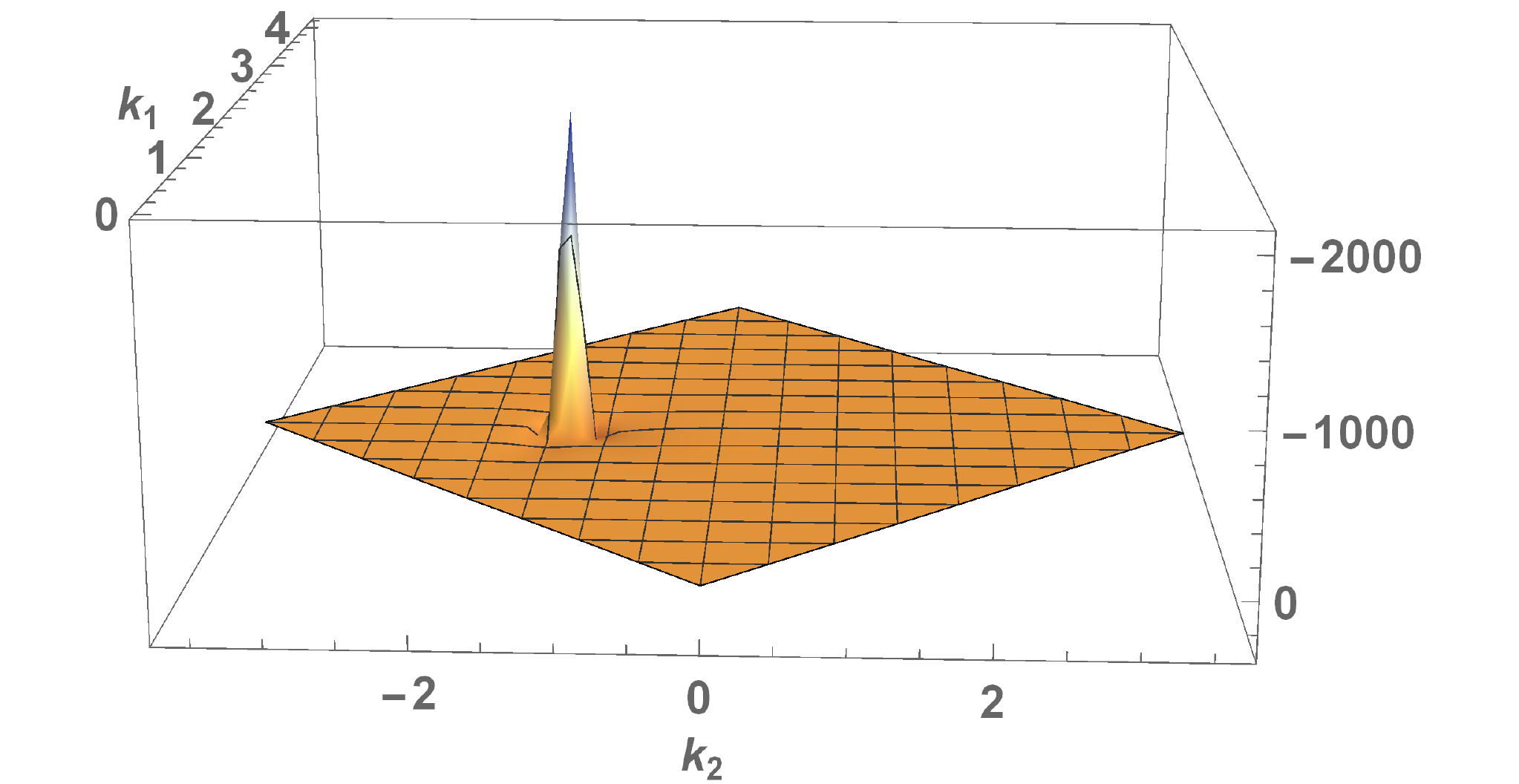}
\caption{}	\label{fig_1d_res}
\end{subfigure}
\cc
\caption{(Color online){We consider the Haldane model subjected to a linear quench for a $40\times 40$ lattice with, $\tau=1600$. The quench parameters are $t_1=2.0$, $t_2=1.0$,
$\phi=-\pi/2$, $M_i=3\sqrt{3}t_2+2.5$, $M_f=3\sqrt{3}t_2-2.5$.  (a)-(c) The angle $\theta$ of the instantaneous eigenstate at all points of the Brillouin zone (BZ) 
(see Appendix \ref{sec_haldane}) for $t=0$, $t=t_c=\tau/2$ and $t=\tau$, respectively for a linear adiabatic quench ($\tau\simeq\tau_a$) from the trivial phase to the topological phase. (d) The curvature at $t=t_c$ shows a monopole singularity due to the instanton in $(k_x,k_y,t_c)$ at the Dirac point.}}
 \end{figure*}

\subsection* {$U(1)$ monopole and topological charge}
\label{monopole_app}
	Here, we have chosen the instantaneous eigenstates $\ket{\phi_{k\pm}(t)}$ of the instantaneous Hamiltonian $H_k(t)=\vec{h}(k,t).\vec{\sigma}$ as,
	\begin{widetext}
		\begin{equation}\label{eq:igs}
	\ket{\phi_{k-}(t)}=\begin{pmatrix} -e^{-i\phi}\sin{\frac{\theta}{2}} \\ \cos{\frac{\theta}{2}} \end{pmatrix},
	\ket{\phi_{k+}(t)}=\begin{pmatrix} e^{-i\phi}\cos{\frac{\theta}{2}} \\ \sin{\frac{\theta}{2}}, \end{pmatrix}
	\end{equation}
	where,
	
	\begin{equation}
	\theta(k,t)=\arccos\left(\frac{h_z(k,t)}{|\vec{h}(k,t)|}\right),~~\text{and}~~\phi(k,t)=\arctan\left(\frac{h_y(k,t)}{h_x(k,t)}\right)
	\end{equation}
	\end{widetext}
	When adiabaticity is maintained, the instantaneous ground state of the system is almost occupied while the instantaneous excited states are vacant at all times. Thus, $n_{k-}(t)\simeq 1$ while $n_{k+}(t)\simeq 0$ for all $k$ (in large but finite size systems). Therefore, for an adiabatic situation, the averaged connection $\tilde{A}$, essentially reduces to that of the instantaneous ground state (as also described in the manuscript) and the monopole charge is that of the lower energy band $\ket{\phi_{k-}(t)}$, which in turn is integer quantised for large system sizes. Hence, when the instantaneous Hamiltonian crosses the critical point (say at $t=t_c$), the instanton appears in the gauge curvature (or the $U(1)$ field) as a singularity (Fig.~\ref{fig_1d_res}) at the Dirac point. This is in exact correspondence with the field of a magnetic monopole. {In Fig.~\ref{fig_1b} of the main text, we show that the net flux of this charge when integrated over the complete BZ precisely gives the topological charge according to Gauss theorem. This is simply the well-established Chern number of a system.}
	
	 However, the topological charge is not integer quantised for a generic non-adiabatic dynamics as the band averaged quantity $\tilde{A}$ is no longer a $U(1)$ gauge connection (as has also been discussed in the main text).\\
	
	The topological transition can also be understood as an emergent obstruction in defining an universal gauge in 2+1 D ($k_x$, $k_y$, $t$). It is established that the gauge connection around a $U(1)$ monopole cannot be uniformly defined within a consistent universal gauge choice. This is reflected in a line singularity in the potential originating at the monopole and extending to infinity, also known as the Dirac string. 
	
	Now, as evident from Eq.~\eqref{eq:igs}, this specific gauge choice for  $\ket{\phi_{k-}}$ used, works well for all $\theta$ except at the south pole $\theta=\pi$ of the Bloch sphere. What happens in a trivial phase $(t<t_c)$ is that $\forall~(k_x,k_y)$, the vector $\ket{\phi_{k-}(t)}$ in Eq.~\eqref{eq:igs} is confined in a region which can be described in a smooth uniform gauge (see Fig.~\ref{fig_1a_res}-\ref{fig_1c_res}). However, for ($t>t_c$) the vector $\ket{\phi_{k-}(t)}$ reaches the south pole where the gauge choice fails.\\
	
\subsection*{Gauge invariance}\label{ap:gauge_invariance}
We observe that the defined quantity $\mathcal{C}^U$ is gauge invariant owing to the non-interacting nature of the problem,

\begin{equation}\label{eq:beta_res}
\mathcal{C^U}\propto\int dk_1\partial_{k_1}\beta(k_1),
\end{equation}
upto gauge invariant constants and where,
\begin{equation}
\beta(k_1)=\sum\limits_{\alpha}\int dk_2 n_{\alpha}^kA^{\alpha}_2.
\end{equation}
Expanding Eq.~\eqref{eq:beta_res}, one finds,
\begin{widetext}
\begin{equation}\label{eq:expand_res}
\int dk_1\partial_{k_1}\beta(k_1)=\sum\limits_{\alpha}\int dk_1dk_2\left(n^k_{\alpha}\partial_{k_1}A^{\alpha}_2+A^{\alpha}_2\partial_{k_1}n^k_{\alpha}\right)={\rm I}+{\rm II}
\end{equation}
\end{widetext}
The first term ${\rm I}$ is the intrinsic Hall conductivity after the removal of interband coherences and also manifestly gauge invariant. We therefore proceed to study the transformation of the second term under a local $U(1)$ gauge transformation.

Under a local $U(1)$ gauge transformation, $\ket{\phi^k_{\alpha}}\rightarrow e^{i\gamma(k)}\ket{\phi^k_{\alpha}}$, the populations $n^k_{\alpha}$ being expectations of Hermitian operators, remain manifestly gauge invariant. However, the quantities $A^{\alpha}_{2}$ transform as,
\begin{equation}
A^{\alpha}_{2}\rightarrow A^{\alpha}_{2}+i\partial_{k_2}\gamma
\end{equation}
Therefore, the second term ${\rm II}$ in Eq.~\eqref{eq:expand_res}, under the gauge transformation gain an additional term of the form,
\begin{widetext}
\begin{equation}\label{eq:gauge1_res}
{\rm II}\rightarrow{\rm II}+i\int dk_1dk_2\left(\partial_{k_2}\gamma\right)\left(\partial_{k_1}\sum\limits_{\alpha}n^k_{\alpha}\right)= {\rm II}+i\int dk_1dk_2\left(\partial_{k_2}\gamma\right)\left(\partial_{k_1}N(k)\right),
\end{equation}
\end{widetext}
where $N(k)$ is the expectation of the number operator for each $k$-mode.
However, since the $k$-modes do not scatter among each other and remain decoupled from each other, the total occupation $N(k)$ of each $k$-mode when summed over all the single-particle bands must be constant and depend only on their equilibrium filling. Starting from an equilibrium half-filled state ($N(k)=1~\forall~k$), the second term in the RHS of Eq.~\eqref{eq:gauge1_res} vanishes. This makes the quantity defined in Eq.~\eqref{eq:beta_res} invariant under local $U(1)$ transformations.\\

\section{A brief review on Haldane model of graphene :} 

\label{sec_haldane}

The bare Hamiltonian for the Haldane model  \ct{haldane83} is obtained by breaking the time reversal and sublattice of graphene ,
\begin{widetext}
\begin{equation}
H_{\alpha,\beta,n,m}^0=-t_1\sum\limits_{\left<m\alpha,n\beta\right>}a_{m,\alpha}^{\dagger}a_{n,\beta}+M\sum\limits_{n} a_{n,A}^{\dagger}a_{n,A}
-M\sum\limits_{n} a_{n,B}^{\dagger}a_{n,B}-\sum\limits_{\left<\left<m\alpha,n\alpha\right>\right>}t_2e^{i\phi}a_{m,\alpha}^{\dagger}a_{n,\alpha}+h.c.,
\end{equation}
\end{widetext}	
where the real nearest neighbour (N1) hopping $t_1$ (with $t_2=0,M=0$) comprises the bare graphene Hamiltonian; the indices $n$ and $\alpha$ represent site and sublattice respectively. The diagonal staggered mass (Semenoff mass) $M$ explicitly breaks the sublattice symmetry of the model. Further the complex nest nearest neighbour (N2) hopping term $t_2$, is applied such that the time reversal symmetry is broken in the next nearest neighbour hopping while the net flux through each plaquette remains zero.
The Haldane model is known to exhibit non-trivial Chern topology when its ground state is completely filled depending on the parameters $M$, $t_1$, $t_2$ and $\phi$.

\begin{figure}[h]
\centering
\includegraphics[width=.45\textwidth,height=5.cm]{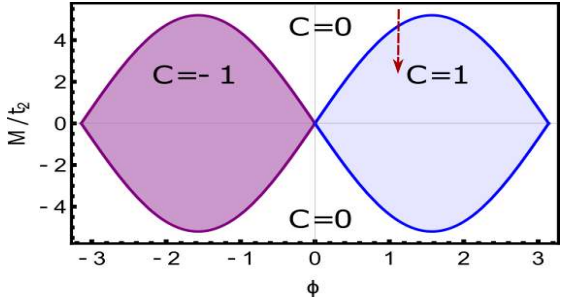}
\caption {(Color online) The topological phase diagram of the Haldane model with $t_1=1.0$. The distinct topological phases are separated by quantum critical lines on which the parameter values are such that the system becomes gapless. The parameter regions showing non-zero values of the Chern number (C) are topologically non-trivial. The red arrow show the direction of the initial and final region of a quench from a trivial phase to a topological phase.}	
\label{fig:phase_diagram_ap}
\end{figure}

Interestingly, the Haldane model with explicitly broken time reversal symmetry is known to host topologically non-trivial phases for certain parameter regions. The topology of the Hamiltonian is essentially the homotopy classification of the map $(k_1,k_2)\rightarrow H^k(k_1,k_2)$ in reciprocal space and is characterised by the gauge invariant Chern topological invariant,
\begin{equation}
C=\frac{1}{\left(2\pi\right)^2}\int_{BZ}dk_1dk_2\mathcal{F}_{12}(\ket{\psi_k}),
\end{equation}
where, $\mathcal{F}_{12}(\ket{\psi_k})$ is the $U(1)$ curvature defined over the ground state $\ket{\psi_k}$ of the Hamiltonian $H^k$, i.e.,
\begin{equation}
\begin{split}
\mathcal{F}_{12}(\ket{\psi_k})= \partial_{k_2}\langle{\psi_k|\partial_{k_1}|\psi_k}\rangle-\partial_{k_1}\langle{\psi_k|\partial_{k_2}|\psi_k\rangle}.
\end{split}
\end{equation}
The Chern invariant is integer quantized as long as the Hamiltonian $H^k$ does not approach a QCP where the Chern number becomes ill-defined. Different integer values of the Chern number characterize distinct topological phases separated by QCPs (see Fig.~\ref{fig:phase_diagram_ap}). 

{Each point on the Bravias lattice can be referenced in terms of the Bravias lattice vectors, i.e.,
	\begin{equation}
	\vec{a}=n_1\vec{a}_1+n_2\vec{a}_2,
	\end{equation}
	where the vectors $\vec{a}_1$ and $\vec{a}_2$ span the Bravias lattice and $n_1,n_2$ are integers.
	We choose the vectors $\vec{a}_1$ and $\vec{a}_2$ to be the next nearest neighbour hopping vectors such that,
	\begin{equation}\label{eq:lattice:ap}
	\begin{split}
	\vec{a}_1=\vec{\Delta}_{22},\\
	\vec{a}_2=-\vec{\Delta}_{21},
	\end{split}
	\end{equation}
	
	\begin{figure*}
	\begin{subfigure}{0.45\textwidth}
	\includegraphics[width=\textwidth]{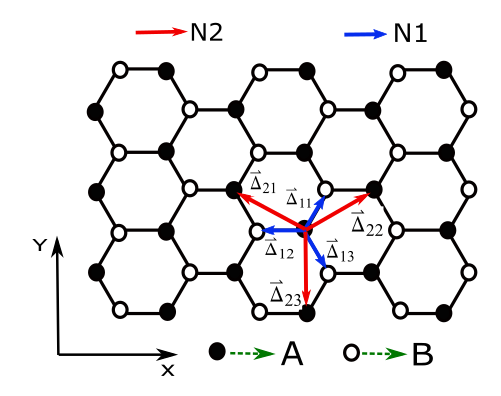}
	\caption{} \label{fig_1a_ap} 
	\end{subfigure}
	\quad\quad\begin{subfigure}{0.45\textwidth}
	\includegraphics[width=\textwidth]{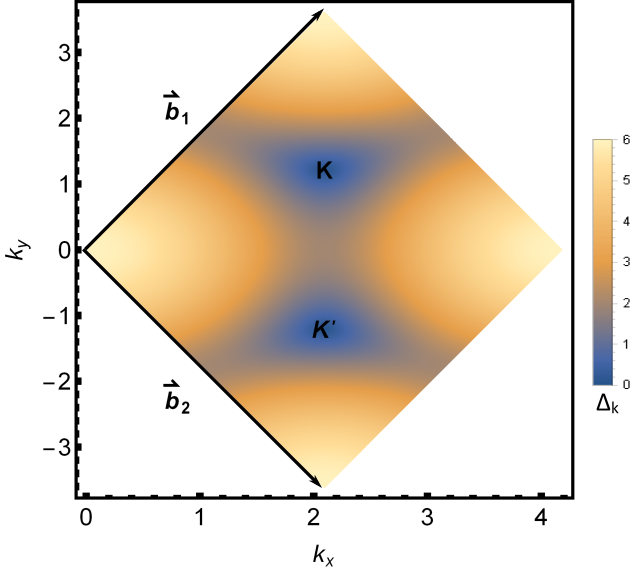}
	\caption{}	\label{fig_1b_ap}
	\end{subfigure}
	\caption{(Color online) (a) The hexagonal graphene lattice showing the nearest neighbour (N1) and next-nearest neighbour (N2) hopping vectors $\vec{\Delta}_{1i}$ and $\vec{\Delta}_{2i}$, respectively, where the lattice constant is set to be $a=1$. The hollow and the filled atoms represent the B and A sublattices respectively. (b) The Brillouin zone of graphene spanned by the reciprocal lattice vectors $\vec{b}_1$ and $\vec{b}_2$ containing two inequivalent Dirac points $K$ and $K^{\prime}$ (the cartesian directions has been labelled by $k_x$ and $k_y$ respectively). The color density shows the absolute value of the bandgap 
		$\Delta_k$ of the reciprocal space graphene Hamiltonian showing vanishing gaps at the Dirac points for a $600\times 600$ lattice size having the N1 hopping strength $t_1=1.0$ and the N2 hopping $t_2=0$.}
	\end{figure*}

	where $\vec{\Delta}_{2i}$ are the $N2$ vectors as shown in Fig.~\ref{fig_1a_ap}.
	
	Invoking the discrete translational invariance of the Hamiltonian one can employ a discrete Fourier transform to decouple the Hamiltonian $H(t)$ in momentum space. The reciprocal space is spanned by the reciprocal lattice vectors $\vec{b}_1$ and $\vec{b}_2$, i.e. every reciprocal lattice point can be represented as,
	\begin{equation}
	\vec{b}=k_1\vec{b}_1+k_2\vec{b}_2,
	\end{equation}
	where, $k_1$, $k_2\in[0,1)$. We choose a rhomboidal Brillouin zone spanned by reciprocal lattice vectors $\vec{b}_1$ and $\vec{b}_2$ (see Fig.~\ref{fig_1b_ap}) containing two independent Dirac points $K$ and $K^{\prime}$}. In our choice of representation,
\begin{equation}
\vec{b}_1=\frac{2\pi}{3a}\{1,\sqrt{3}\}~~\text{and}~~\vec{b}_2=\frac{2\pi}{3a}\{1,-\sqrt{3}\},
\end{equation}
where we have chosen $a=1$. The corresponding inequivalent Dirac points in the Brillouin zone shown in Fig.~\ref{fig_1b_ap} are given by,
\begin{equation}
K=\frac{2\pi}{3}\left(1,\frac{1}{\sqrt{3}}\right)~~\text{and}~~K^{\prime}=\frac{2\pi}{3}\left(1,-\frac{1}{\sqrt{3}}\right).
\end{equation}

The bare Haldane Hamiltonian gets decoupled in the momentum space where $H^0(k)$ can be written in the basis $\ket{k,A}$ and $\ket{k,B}$ as,
\begin{equation}
H^0(k)=\vec{h}(k).\vec{\sigma} =h_x(k) \sigma_x+h_y(k) \sigma_y+h_z(k) \sigma_z,
\label{eq_hamil_k_ap}
\end{equation}
such that,
\begin{equation}\label{eq:bloch_ham_ap}
\begin{split}
h_x(k)=-t_1\sum\limits_{i=1}^{3}\cos{\left(\vec{k}.\vec{\Delta}_{1i}\right)},\\
h_y(k)=-t_1\sum\limits_{i=1}^{3}\sin{\left(\vec{k}.\vec{\Delta}_{1i}\right)},\\
h_z(k)=M-t_2\sin{\phi}\sum\limits_{i=1}^{3}\sin{\left(\vec{k}.\vec{\Delta}_{2i}\right)},
\end{split}
\end{equation}
$\vec{\Delta}_{1i}$ and $\vec{\Delta}_{2i}$ are the nearest neighbour and next nearest neighbour lattice vectors respectively (see Fig.~\ref{fig_1a_ap}) chosen to be,
\begin{equation}
\begin{split}
\vec{\Delta}_{11}=\frac{a}{2}\{1,\sqrt{3}\},~~\vec{\Delta}_{12}=\{-a,0\},~~\vec{\Delta}_{13}=\frac{a}{2}\{1,-\sqrt{3}\} ~~\text{and},\\
\vec{\Delta}_{21}=\frac{a}{2}\{-3,\sqrt{3}\},~~\vec{\Delta}_{22}=\frac{a}{2}\{3,\sqrt{3}\},~~\vec{\Delta}_{23}=\{0,-a\sqrt{3}\},
\end{split}
\end{equation}
in the cartesian frame (Fig.~\ref{fig_1a_ap}) where we have chosen the lattice parameter $a=1$.
Note that we have used Eq. \eqref{eq_hamil_k_ap} where  the Semenoff mass term which  appears only in $h_z(k)$ 
is linearly ramped across the quantum critical point from the non-topological to the topological phase.

	\section{Currents}
	
	\label{App_current}
	
	The definition of the topological classification of out of equilibrium states is directly connected to the evolution of particle currents generated in the time dependent state of the system. {For the topological invariant to conform with the adiabatic edge current dynamics, it is essential to take note of the time evolution of the current operator. 
		
		%
		%
		%
		%
		%
		%

		As discussed in Refs. \ct{rigol15,marin14}, the measured particle current in the out of equilibrium system is dependent on the instantaneous Hamiltonian.
		This can be easily seen by explicitly computing the expectation of the current operator between two sites when the system is driven out of equilibrium.
		%
		Referring to the Haldane Hamiltonian and resorting to the Heisenberg picture,
		\begin{equation}\label{eq:pop_dyn_ap}
		\frac{d(a_n^{\dagger}a_m)}{dt}=-i\left[H(t),a_n^{\dagger}a_m\right].
		\end{equation}
		As the dynamics is unitary, the mean rate of change of local population at a site is directly proportional to the average local current at the site. Thus, the expectation,
		\begin{equation}\label{eq:pop_curr_ap}
		\left<\frac{d(a_n^{\dagger}a_m)}{dt}\right>=\sum\limits_{n}J_{nm},
		\end{equation}
		where $J_{nm}$ is the average current between the sites $i$ and $j$. Comparing Eq.~\eqref{eq:pop_dyn_ap} and Eq.~\eqref{eq:pop_curr_ap}, one obtains,
		\begin{equation}
		J_{nm}={\rm Im}\left[2H_{nm}(t)\left<a_n^{\dagger}a_m\right>\right],
		\end{equation}
		where $H_{nm}(t)$ is the single particle time dependent Hamiltonian,
		\begin{equation}
		\begin{split}
		H(t) =\sum\limits_{nm}H(t)_{nm}a_{n}^{\dagger}a_{m},
		\end{split}
		\end{equation}.
		To evaluate the edge currents we impose semi-periodic boundary conditions on the 2D lattice. Generically, as defined above, the single particle current can be decomposed as,
		\begin{equation}\label{curr_ap}
		\braket{\vec{J}_{SS}}=\braket{\vec{J}_N}+\braket{\vec{J}_{NN}},
		\end{equation}
		where $\vec{J}_{N}$ and $\vec{J}_{NN}$ are the nearest neighbour and the next nearest neighbour current operators respectively,
		\begin{equation}
		\begin{split}
		\braket{J_{Nn}^x}=\sum\limits_{m}t_1\braket{a_{n}^{\dagger}a_m}-hc \\
		\braket{J_{NNn}^x}=\sum\limits_{m}t_2\braket{a_{n}^{\dagger}a_m}-hc,
		\end{split}
		\end{equation}
		where $\braket{J_{Nn}^x}$($\braket{J_{NNn}^x}$) is the nearest(next nearest) current at the $n^{th}$ site and the summation indices extends over all nearest (nest-nearest) neighbour sites to the $n^{th}$ site. Considering the lattice to be periodically wrapped in the x-direction (see Fig.~\ref{fig_1a_ap}) while being open in the y-direction, one obtains two arm-chair edges at the ends of the cylinder. We compute the total horizontal current flowing in the periodic x-direction on one of the arm chair edges $J^x_L$ for a $L\times L$ lattice, in the post quench state to re-establish the bulk boundary correspondence which is depicted in Fig.~\ref{fig_5} of the main manuscript.\\

		\section{ Numerical and experimental generation of the counter-diabatic mass}
		
		\label{App_CD}
		The time-dependent generation of the counter-diabatic term in Eq.~\eqref{eq:unitary_CD1} of the main text can be realised experimentally by a temporal modulation of the nearest neighbour hopping amplitude along a particular direction in the real lattice. This is experimentally realised by the application of small anisotropic strain on the graphene lattice. {The application of strain changes the $C-C$ bond length between and thus renormalises tunnelling amplitudes anisotropically. It is established that under a strain, the hopping energies are modified as} \ct{peeters16},
		{\begin{equation}
			t_{ij}\sim t_0e^{-\beta\left(\frac{l_{ij}}{a_0}-1\right)},
			\end{equation}
			where $l_{ij}$ are bond lengths under strain while $a_0$ is the nearest neighbour bond length (see Sec. 1) {that is proportional to the nearest neighbor hopping $t_0$} in unstrained graphene and $\beta\sim3.37$ is a dimensionless modulation factor. Therefore, with the application of small anisotropic strain, each nearest neighbour hopping strength can be differentially modulated to generate an effective pseudo-magnetic field in graphene which is the essence of the CD protocol.}

		In Fig.~\ref{fig_5} of the manuscript we explicitly demonstrate this by applying a time dependent modulation to the hopping strength along the direction $\vec{\Delta}_{12}$ while keeping the other two nearest-neighbour and next-nearest neighbour hopping unaffected,
		\begin{equation}\label{eq:CD_sup}
		\begin{split}
		t_{\vec{\Delta}_{12}}(t)=-t_1-G\sin{\left(\frac{\pi t}{\tau}\right)};~~G \ge 0\\
		t_{\vec{\Delta}_{11}}=-t_1,\\
		t_{\vec{\Delta}_{13}}=-t_1,
		\end{split}
		\end{equation}

		for the duration of the quench, i.e. $t\in[0,\tau]$ (see Fig.~\eqref{fig_5} of main manuscript) and {$G$ represents the anisotropic strain}. 
		{Note that the term $G \times \sin\left(\frac{\pi t}{\tau}\right)$ term vanishes both 
			at $t=0$ and $t=\tau$ but not at the quantum critical point thus maintaining a finite gap at the topological  critical point of unstrained graphene. For the translationally invariant situation (periodic boundary condition), this term can be shown to modify the $h_x(k)$ of the reduced $2\times2$ Hamiltonian} (see Eq.~\eqref{eq_hamil_k_ap}). Under semi-open boundary conditions, we analyse the entire
		real space Hamiltonian along with the counter-diabatic  term to evaluate the post-quench real time edge current in Fig.~\ref{fig_5} of the main text.
		
		Such anisotropic modulations can be generated experimentally by applying anisotropic strain on the graphene lattice and then modifying it temporally to open up a controlled gap in the spectrum \ct{neto09,peeters13} which in turn suppresses diabatic excitations while crossing a quantum critical point. {Eventually,
			at the final  time $t=\tau$, the anisotropic strain is removed and the lattice returns to its unstrained form provided the maximum applied strain is within the elastic limit of graphene.\\}

		}
		
		\section{Edge Current and CD protocol}
		
		\label{App_edge}
		\begin{figure*}
		\includegraphics[width=0.5\textwidth,height=0.25\textheight]{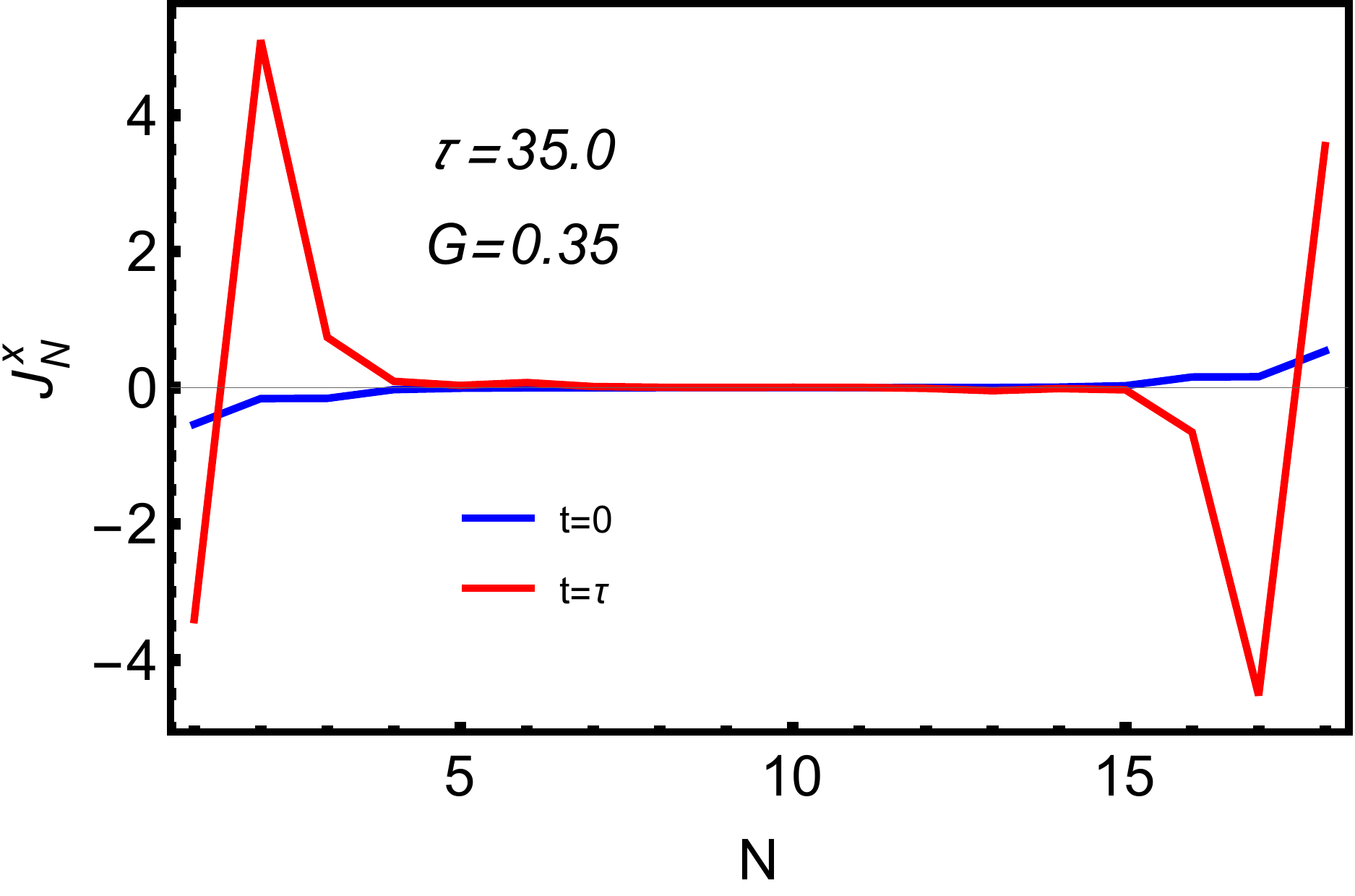}
		\caption{Localisation of the single-particle edge current generated in the initial and the post-quench Haldane model with a cylindrical geometry. The index $N$ denotes the position of a strip along the translational symmetry-broken direction of the cylinder. The post-quench current is observed to be localised into the boundary of the lattice. The simulation is performed for a $18\times 18$ lattice. All the other quench parameters are exactly similar to that used in the paper.} \label{fig:edge_sup} 
		\end{figure*}
		
		 {To establish the emergent topological nature of the post-quench state, we have explicitly checked the boundary localisation of the generated edge currents as defined in Eq.~\eqref{curr}. It is straight-forward to see that the eigenstates of the post-quench Hamiltonian is expected to host conducting edge states. However, the dynamical population of the edge states of the final Hamiltonian is in itself an emergent phenomena which has been demonstrated in the manuscript. Furthermore, it is essential that the edge-states lie in the bulk gap for the edge currents to get boundary-localised. This is ensured by half-filling in the equilibrium system and by suppressing diabatic excitations in the out-of equilibrium state while at the same time, populating the edge channels. It is in this aspect that we discuss it as an indicator of the post-quench system comprising of the edge states and the bulk.
			To exemplify this, we plot the chiral current in a cylindrical geometry in strips along the periodic direction in Fig.~\ref{fig:edge_sup}. We show that the CD post-quench current is well-localised in the edges and decay rapidly into the bulk. The dynamical behaviour of the edge current is similar even without the CD term, nevertheless, the latter facilitates a quicker preparation
			as discussed before.
		}

\end{document}